\documentclass{ws-jai}
\usepackage[flushleft]{threeparttable}
\usepackage{amsmath,amsfonts,amssymb}
\usepackage{graphicx}
\usepackage{setspace}
\usepackage{tocloft}
\usepackage{siunitx}
\usepackage{subcaption}
\usepackage{caption}
\usepackage{soul}
\usepackage[normalem]{ulem}

\begin{document}

\catchline{}{}{}{}{} % Publisher's Area please ignore

\markboth{B.D. Donovan et al.}{Performance Testing of a Large-Format Reflection Grating Prototype for a Suborbital Rocket Payload}

\title{Performance Testing of a Large-Format X-ray Reflection Grating Prototype \\ for a Suborbital Rocket Payload}

\author{Benjamin D. Donovan$^{1,4}$, Randall L. McEntaffer$^{1}$, Casey T. DeRoo$^{2}$, James H. Tutt$^{1}$,\\  Fabien Gris{\'e}$^{1}$, Chad M. Eichfeld$^{1}$, Oren Z. Gall$^{1}$, Vadim Burwitz$^{3}$, Gisela Hartner$^{3}$,  \\Carlo Pelliciari$^{3}$, and Marlis-Madeleine La Caria$^{3}$ \\ \vspace{5pt}}

\address{
$^{1}$The Pennsylvania State University, University Park, PA 16802, USA\\
$^{2}$The University of Iowa, Iowa City, IA 52242, USA\\
$^{3}$PANTER X-ray Test Facility, Max Planck Institute for Extraterrestrial Physics, \\ 82061 Neuried, Germany\\
$^{4}$bdonovan@psu.edu
}

\maketitle

\corres{$^{4}$Corresponding author.}

\begin{history}
\received{(to be inserted by publisher)};
\revised{(to be inserted by publisher)};
\accepted{(to be inserted by publisher)};
\end{history}

% --- OLD ----
\begin{abstract}
The soft X-ray grating spectrometer on board the Off-plane Grating Rocket Experiment (OGRE) hopes to achieve the highest resolution soft X-ray spectrum of an astrophysical object when it is launched via suborbital rocket. Paramount to the success of the spectrometer are the performance of the $>250$ reflection gratings populating its reflection grating assembly. To test current grating fabrication capabilities, a grating prototype for the payload was fabricated via electron-beam lithography at The Pennsylvania State University's Materials Research Institute and was subsequently tested for performance at Max Planck Institute for Extraterrestrial Physics' PANTER X-ray Test Facility. Bayesian modeling of the resulting data via Markov chain Monte Carlo (MCMC) sampling indicated that the grating achieved the OGRE single-grating resolution requirement of $R_{g}(\lambda/\Delta\lambda)>4500$ at the 94\% confidence level. The resulting $R_g$ posterior probability distribution suggests that this confidence level is likely a conservative estimate though, since only a finite $R_g$ parameter space was sampled and the model could not constrain the upper bound of $R_g$ to less than infinity. Raytrace simulations of the system found that the observed data can be reproduced with a grating performing at $R_g=\infty$. It is therefore postulated that the behavior of the obtained $R_g$ posterior probability distribution can be explained by a finite measurement limit of the system and not a finite limit on $R_g$. Implications of these results and improvements to the test setup are discussed.
\end{abstract}


\section{Introduction}
\noindent The soft X-ray bandpass (0.3 -- 2.0 keV) hosts a variety of astrophysically abundant absorption and emission lines. These transition lines provide a wealth of information about astrophysical plasmas and other highly-energetic astrophysical phenomena throughout the Universe \citep{Brenneman:2016}. The only instrument capable of probing these transition lines at high resolution ($R[\lambda/\Delta\lambda]>2000$) is a grating spectrometer. Current grating spectrometers onboard the \textit{Chandra X-ray Observatory} \citep{Weisskopf:1988aa} and \textit{XMM-Newton} \citep{Jansen:2001aa} can achieve modest resolution, but technological advances since the launch of these missions offer the next generation of spectrometers significantly higher performance to explore the soft X-ray transition lines in more detail. One such technology which can offer this higher performance is an X-ray reflection grating operated in the extreme off-plane mount \citep{McEntaffer:2013aa}. This technology will be implemented on an upcoming soft X-ray spectroscopy sub-orbital rocket payload -- the Off-plane Grating Rocket Experiment (OGRE) \citep{Donovan:2019aa}. This payload will use state-of-the-art reflection gratings manufactured using electron-beam (e-beam) lithography at the Materials Research Institute at The Pennsylvania State University; however, these gratings first need to be tested to evaluate their performance and to determine if they can meet OGRE performance goals.

\begin{figure*}
    \begin{subfigure}[ht]{0.48\linewidth}
        \includegraphics[width=\linewidth]{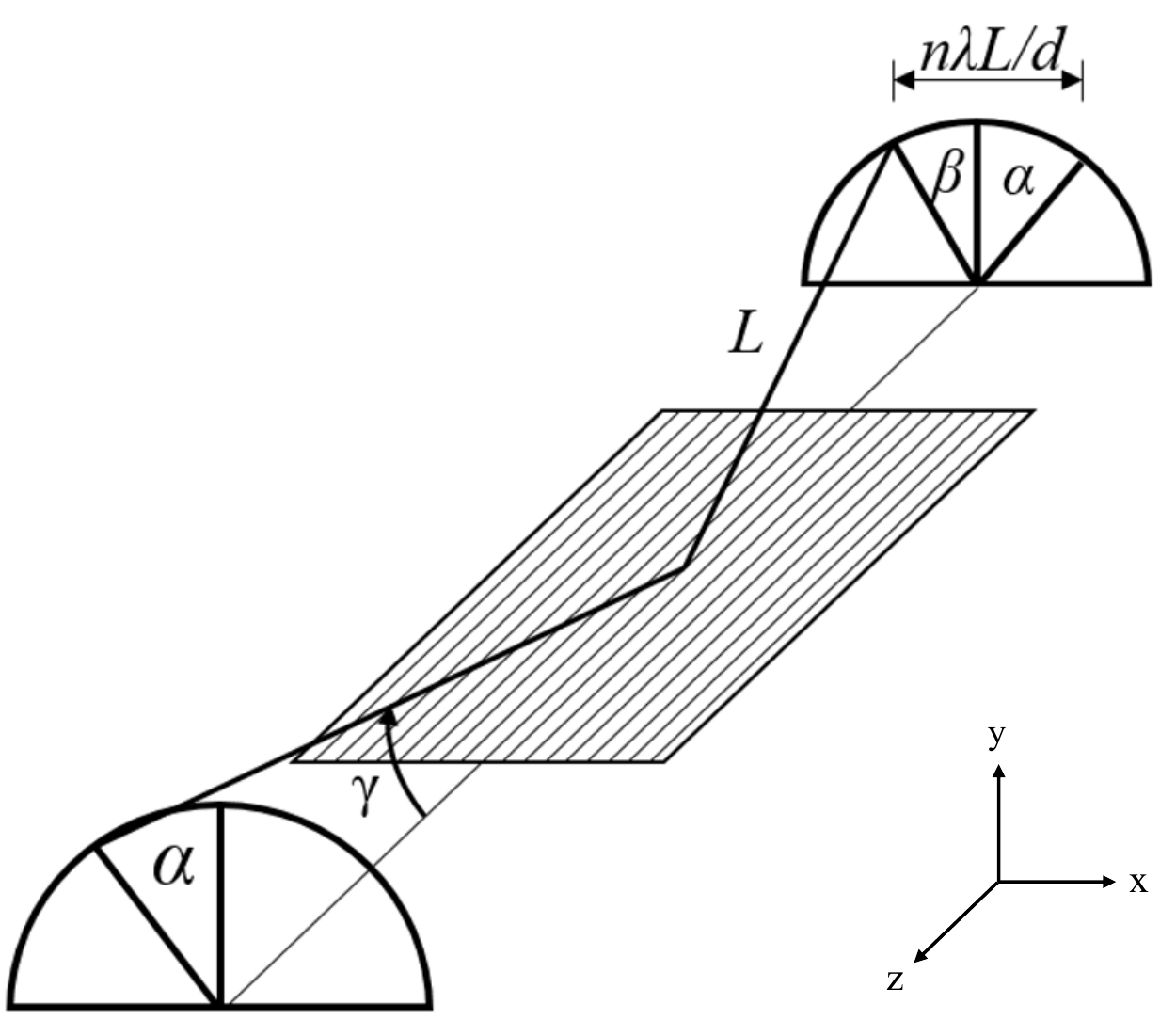}
        \caption{}
    \end{subfigure}
    \hfill
    \begin{subfigure}[ht]{0.48\linewidth}
        \includegraphics[width=\linewidth]{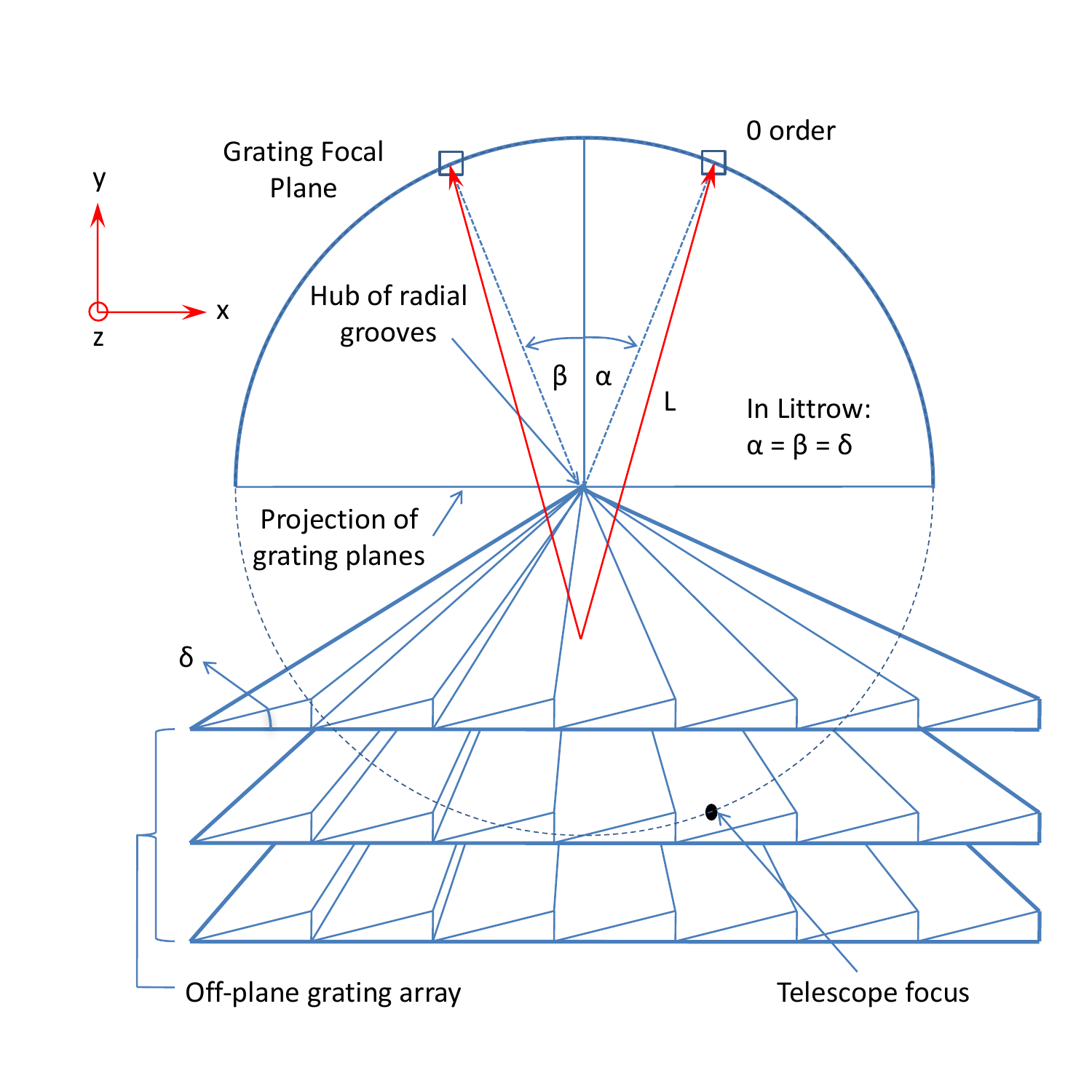}
        \caption{}
    \end{subfigure}
\caption{The diffraction geometry of reflection gratings operated in the extreme off-plane mount. (Left) Light is incident on the grating quasi-parallel to the groove direction and is directed into an arc on the focal plane following the generalized grating equation described in Eq. 1. (Right) An idealized array of three aligned gratings with radial grooves for increased spectral resolution and with facet angle $\delta$ in the Littrow mount ($\alpha=\beta=\delta$) for maximal diffraction efficiency. Figure reprinted from \citet{McEntaffer:2013aa}.}
\label{fig:grating_geometry1}
\end{figure*}

\subsection{Diffraction in the Extreme Off-plane Mount}

The diffraction geometry of a reflection grating operated in the extreme off-plane mount can be seen in left-hand side of Figure \ref{fig:grating_geometry1} and follows the generalized grating equation:

\begin{equation}
\sin{\alpha}+\sin{\beta}=\frac{n\lambda}{d\sin{\gamma}},
\label{eq:1}
\end{equation}

\noindent where $\gamma$ is the polar angle of the incident X-rays with s wavelength of $\lambda$ defined from the groove axis at the intersection point, $d$ is the line spacing of the grooves, $n$ is the diffraction order, $\alpha$ is the azimuthal angle along a cone with half-angle $\gamma$, and $\beta$ is the azimuthal angle of the diffracted light \citep{Cash:1983aa}. In the extreme off-plane mount, photons intersect the grating at grazing incidence and quasi-parallel to the groove direction ($\gamma\lesssim2^{\:\circ}$). Photons are then diffracted into an arc on the focal plane with radius $R=L\sin\gamma$, where $L$ is the distance the photon travels from the grating to the focal plane. While diffraction is confined to an arc, the spectral information is contained only in the axis formed by the intersection of the focal plane and the grating plane (the dispersion direction): 

\begin{equation}
\frac{d\lambda}{dx}=\frac{10^{7}}{n\text{L}\text{D}}\frac{\mathring{A}}{\text{mm}},
\label{eq:2}
\end{equation}

\noindent where $D$ is the groove density ($\equiv1/d$) and $\hat{x}$ is the dispersion direction with $x\equiv L\sin\gamma(\sin\beta+\sin\alpha)$.

Reflection gratings in the extreme off-plane mount are typically paired with an X-ray focusing optic in astronomical applications to form a reflection grating spectrometer. To minimize aberrations seen on the focal plane and therefore increase spectral resolution in this system, these gratings are typically manufactured with grooves that match the convergence of the photons from the X-ray focusing optic. These grooves converge to the center of the diffraction arc as shown in the right-hand side of Figure \ref{fig:grating_geometry1} \citep{Cash:1983aa}. Gratings with this groove pattern are known as ``radially grooved'' gratings. Additionally, reflection gratings can be manufactured to have a blazed profile with facet angle $\delta$. This groove profile enables a grating to preferentially disperse its incident photons to a specific location on the focal plane. The extreme off-plane mount realizes maximum diffraction efficiency in the Littrow configuration ($\alpha=\beta=\delta$) at the blaze wavelength,

\begin{equation}
n\lambda_{b}=2d\sin\gamma\sin\delta,
\end{equation}

\noindent with total diffraction efficiencies approaching the reflectivity of the grating's metallic coating \citep{Miles:2018aa}. The Littrow configuration is also depicted in the right-hand side of Figure \ref{fig:grating_geometry1}. By combining these two properties, reflection gratings operated in the extreme off-plane mount are capable of both high spectral resolution and high diffraction efficiency in the soft X-ray bandpass.

Reflection gratings operated in this diffraction geometry have heritage on several proposed high-resolution soft X-ray spectroscopy missions \citep{Bookbinder:2008aa, Lillie:2011aa, White:2010aa}. In addition to these proposed missions, these reflection gratings will be implemented on the OGRE soft X-ray grating spectrometer \citep{Tutt:2018aa, Donovan:2018bb, Donovan:2019aa}. The OGRE grating spectrometer will observe the binary star system Capella ($\textrm{$\alpha$}$ Aurigae), a bright X-ray source with a line-dominated soft X-ray spectrum. With a performance goal of $R>2000$, the instrument hopes to obtain the highest resolution astronomical spectrum to date in the soft X-ray bandpass. To achieve this goal, the spectrometer will employ high-angular-resolution X-ray optics developed at NASA Goddard Space Flight Center, high-spectral-resolution reflection gratings developed at The Pennsylvania State University, and electron-multiplying CCDs developed by The Open University and XCAM Ltd in the United Kingdom.

\subsection{OGRE Single-Grating Resolution Requirement}
\label{subsect:grat_res_requirement}

Performance errors, misalignments of spectrometer components within the OGRE payload, and pointing errors during flight all conspire to reduce the achievable spectral resolution of the OGRE spectrometer. It is thus imperative that each of these errors is understood so that the instrument can achieve its performance goal of $R>2000$. To determine how each of these errors will affect the performance of the OGRE spectrometer, a raytrace model of the instrument was constructed. Each error was simulated independently with the simulation returning the impact of that particular error on the performance of the spectrometer over a range of potential error values. These simulations were then used to allot realistic requirements to each error or misalignment, while maintaining the overall OGRE performance goal of $R>2000$ \citep{Donovan:2020aa}. 

Included in this comprehensive error budget is the performance requirement of a single grating in the spectrometer: the single-grating resolution requirement. Two error sources conspire to reduce the achievable performance of a single grating: period errors and figure errors. Figure errors arise when photons diffract off a non-flat grating substrate, whereas period errors are the result of a local grating period not matching the ideal period at that location due to errors in the grating's fabrication process. Both of these errors increase the width of the spectral line on the focal plane, thereby decreasing the achievable spectral resolution of the spectrometer. For OGRE, the single-grating resolution requirement is $R_g>4500$. Each grating within the spectrometer must therefore demonstrate a performance of $R_g>4500$ for OGRE to meet its overall performance goal of $R>2000$. 

A single reflection grating can be combined with an X-ray focusing optic and a detector to form a single-grating X-ray spectrometer. The performance of this assembled grating spectrometer can then be tested at an X-ray test facility housing an X-ray generating source. The diffracted-order line-spread function (LSF) formed by this spectrometer in the test facility is the convolution of five main components: the spectral line profile under observation, the finite size of the X-ray source used to generate this spectral line profile, the point-spread function (PSF) of the X-ray optic, slope errors introduced by the figure of the grating substrate, and errors introduced during diffraction due to grating fabrication errors or geometric aberrations. The spectral line profile under observation is generally known from theory or experimental measurements. Thus, if the errors due to the optic, X-ray source size, and grating figure can also be determined, the diffraction-induced errors ($R_g$) can be calculated. In practice, these five components can be determined by examining the optic PSF, the zero-order LSF, and the diffracted-order LSF formed by the spectrometer in the test facility. The optic PSF does not contain any information about grating figure or diffraction-induced errors, and therefore can be used to determine the combined contribution from the optic and X-ray source. If the size of the X-ray source is known, the optic performance can be deconvolved from the X-ray source contribution. Slope errors are introduced into the system during the reflection of the optic PSF off of the grating substrate, and appear in the zero-order LSF of the spectrometer. Therefore, the contribution from the grating figure can be determined by comparing the zero-order LSF to the optic PSF. Finally, optic, X-ray source, and grating figure contributions combine with the spectral line profile and diffraction-induced errors to form the diffracted-order LSF. Thus, the diffraction-induced errors can be calculated since the errors introduced by the optic, X-ray source, and the grating figure have already been determined and the spectral line profile is known from theory or from previous experimental data.

An important distinction should be made between the spectral resolution of the system ($R$) and the grating resolution ($R_g$). The spectral resolution of the system is that realized by the spectrometer as a whole, and represents the ability of the spectrometer to resolve spectral lines. The grating resolution, on the other hand, only represents diffraction-induced errors in the system. This value is used to evaluate the performance of a grating independent of the optic performance, X-ray source size, and grating figure, and thus these contributions are not included in this value. As described above, the ability to constrain $R_g$ is determined by the ability to discern between the five main contributors to the diffracted-order LSF: the optic, the X-ray source size, grating figure, the spectral line profile, and diffraction-induced aberrations. In the case where infinite data are available, the optic performance, X-ray source size, and slope errors induced by the grating substrate do not impact the ability to determine $R_g$, since these contributions are known exactly. However, it is not possible to obtain infinite data in practice. Thus, a trade is typically made between these contributions and statistics to constrain $R_g$. For example, if small amounts of data are available, some combination of an optic with better performance, an X-ray source with a smaller angular extent, and a grating with lower figure errors are required to better constrain $R_g$.

\vspace{11pt}

In this paper, an X-ray test campaign at the Max Planck Institute for Extraterrestrial Physics' PANTER X-ray Test Facility in July 2018 is described. An OGRE grating prototype, manufactured via e-beam lithography at The Pennsylvania State University's Materials Research Institute, was combined with a single-shell X-ray optic and a detector at the focal plane to form a reflection grating spectrometer. The spectrometer operated in an OGRE-like diffraction geometry, and performance was assessed and compared to the single-grating resolution requirement of the OGRE spectrometer. 

\section{Test Apparatus}
\label{sect:testing}

The PANTER X-ray Test Facility hosts a $\sim$120 m long, 1 m diameter beamline. On one end of the beamline is a source chamber which houses an electron impact source with a selection of anodes to produce a variety of X-ray transition lines. At the opposite end of the beamline is a 12 m long, 3.5 m diameter test chamber with vacuum staging to manipulate the components under test. The distance between the source and test chambers serves to effectively collimate the X-ray photons at the test chamber by limiting their allowable divergence angles. 

% For this test, the electron impact source utilized a Mg anode to produce Mg-K$\rm{\alpha}$ photons. The Mg-K$\rm{\alpha}$ line complex is composed of two primary components, Mg-K$\rm{\alpha_{1}}$ and Mg-K$\rm{\alpha_{2}}$, with wavelengths of 0.9889 nm and 0.989 nm respectively in a 2:1 intensity ratio \citep{Schweppe:1994}.

The assembled spectrometer was constructed in the PANTER test chamber and consisted of three main components: a mono-crystalline silicon Wolter I-type \citep{Wolter:1952aa} focusing optic \citep{Zhang:2019aa}, an OGRE reflection grating prototype, and a detector at the focal plane. The quasi-collimated light from the source end of the facility was incident first on the focusing optic. The focusing optic formed a converging X-ray beam which was then intercepted by the reflection grating. The grating diffracted the incident light according to its wavelength, which then continued to the focal plane where it formed a focus on the detector. Each of these three spectrometer components will described in more detail below.

\subsection{Mono-Crystalline Silicon Optic} 
\label{sect:optic}
Mono-crystalline silicon optics are an X-ray optics technology developed at NASA Goddard Space Flight Center \citep{Zhang:2019aa}. The mirrors are produced from a block of single-crystal silicon, which allows them to theoretically be manufactured free of internal stresses for improved mirror performance. Further details of the mono-crystalline silicon mirror manufacture and alignment processes can be found in \citet{Riveros:2019aa} and \citet{Chan:2019aa}. This technology is currently being studied for missions such as \textit{Lynx} \citep{Gaskin:2018aa, Zhang:2019bb}, AXIS \citep{Mushotzky:2018}, HEX-P \citep{Madsen:2019}, STAR-X \citep{McClelland:2017}, and FORCE \citep{Nakazawa:2018}. In addition to these proposed missions, mono-crystalline silicon X-ray optics will also be used on OGRE \citep{Donovan:2019aa}. To date, the technology has produced single-shell Wolter I-type optics with half-power diameters of $<1.3$ arcsec \cite{Zhang:2019aa}. 

% This performance is approaching the performance of \textit{Chandra} (HPD: $\sim0.5$ arcsec) -- the current state-of-the-art in the field -- but does so at a fraction of the mass, allowing for much greater effective areas to be achieved on future missions.

%The mono-crystalline mirror manufacture process is described in detail elsewhere \cite{} and only a brief discussion will be given here. Each mirror begins as a mono-crystalline silicon block. This block is roughly cut to the desired conical shape and then ground/lapped to form the precise conical shape. The surface is then cut from the silicon block with a band saw to the desired thickness, enabling the block to be reused to produce more mirrors. This cut, however, damages the silicon crystal layers on the back surface of the mirror, severely distorting the substrate. To remove these damaged crystal layers and return the mirror figure back to its precise conical shape, the back surface of the mirror is chemically-etched, exposing the low-stress mono-crystalline silicon structure underneath. The mirror is then polished to achieve the desired surface quality and finally the edges are trimmed, yielding the final low-stress, high-resolution X-ray optic segment.

The mono-crystalline silicon optic used for this test campaign was a single-shell Wolter I-type optic with a radius at the paraboloid-hyperboloid intersection node of 156 mm and a focal length of $\sim8.4$ m for a source at infinity. The effective focal length of the optic in the PANTER X-ray Test Facility was $\sim9.1$ m due to the finite distance to the source.  The individual mirror segments of the optic were approximately 100 mm in axial length, 30 degrees in azimuthal extent, and were separated by a 20 mm axial gap. The small azimuthal extent produced a PSF that was much wider in the reflection plane than in the perpendicular plane. 

\begin{figure} 
\centering
\includegraphics[width=0.5\linewidth]{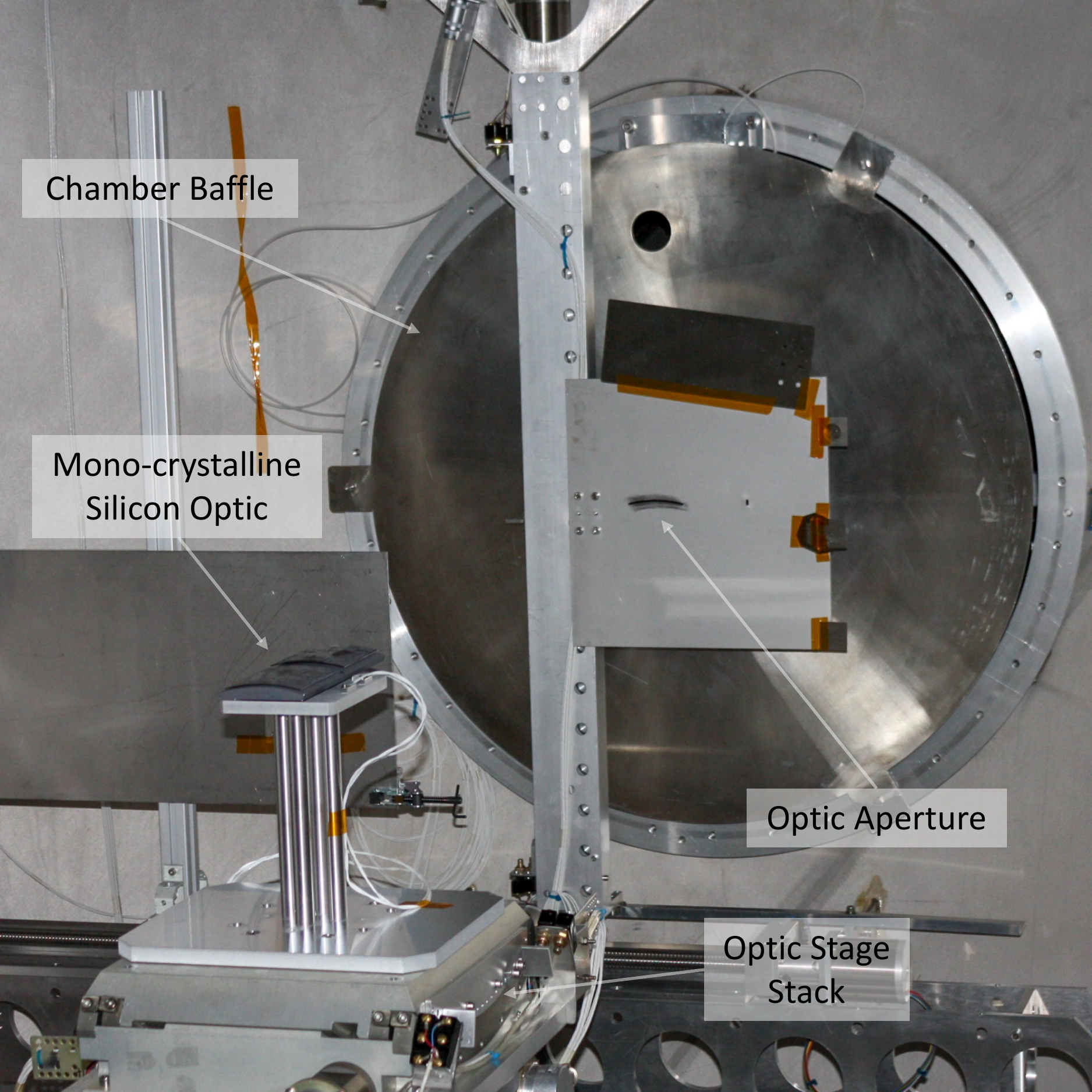}
\caption{The mono-crystalline silicon X-ray optic, its staging, and its aperture implemented in the PANTER test chamber.}
\label{fig:optic_pic}
\end{figure}

The optic was mounted on translation and rotation stages to align the optic to the incident X-ray photons traveling from the source end of the facility. In addition, several apertures were placed into the beam to prevent errant photons from traveling down to the focal plane and to ensure illumination of only the X-ray optic. A picture of the optic, a portion of its associated staging, and its aperture in the test chamber can be seen in Figure \ref{fig:optic_pic}. 

\subsection{Focal Plane}
\begin{figure}
\centering
\includegraphics[width=0.5\linewidth]{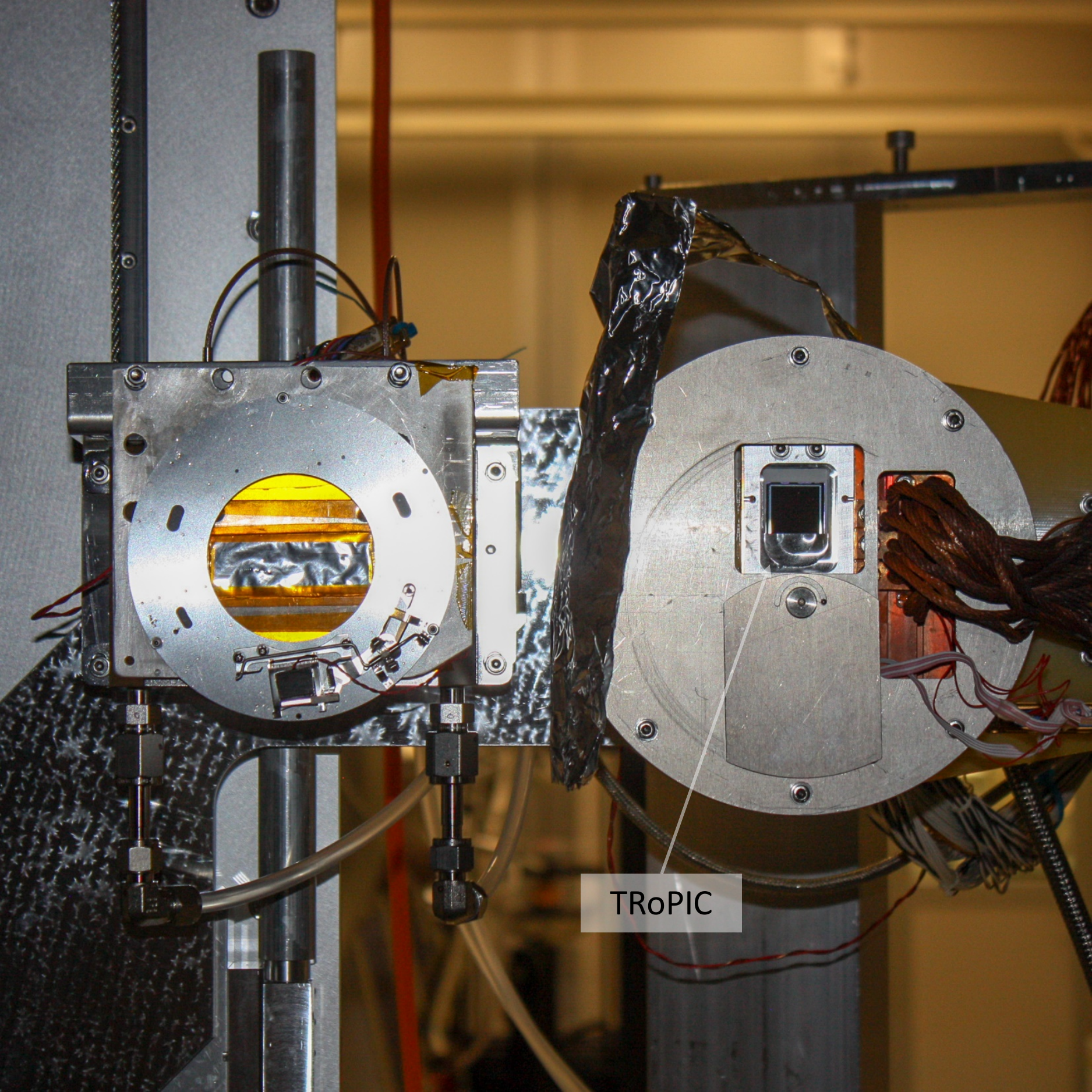}
\caption{TRoPIC implemented in the PANTER test chamber.}
\label{fig:det_pic}
\end{figure}

A single CCD image sensor was used to sample the optic PSF and the diffraction arc produced by the grating. The Third Roentgen Photon Imaging Camera (TRoPIC) is a 256 pixel x 256 pixel device with 75 \si{\micro\metre} pixels and a sufficiently fast frame rate ($20$ Hz) to detect individual photons interacting with the device. The sensor was mounted on three linear stages on the focal plane providing movement in all three translational degrees of freedom. The stage stack was large enough to fully sample the diffraction arc in the grating test geometry (to be described in Section \ref{sect:test_plan}). A close-up image of TRoPIC mounted onto the focal plane stage stack can be seen in Figure \ref{fig:det_pic}.

% TRoPIC is a smaller version of the detector used on eROSITA \citep{Predehl:2010aa}.

\subsection{X-ray Reflection Grating} \label{sub:gratings}
Intercepting the converging beam formed by the mono-crystalline silicon X-ray optic was a single X-ray reflection grating. This grating was the first grating prototype manufactured for the OGRE payload. The groove pattern of the grating had a groove period of 157.57 nm at the center of the grating with radial grooves that converged to a point 3250 mm away. This pattern measured 100 mm in groove direction and $\sim86$ mm in the cross-groove direction (dispersion direction).  

% This groove period decreased closer to the focal plane due to the radial nature of the groove pattern. The center of the grating was designed to be placed 3250 mm from the focal plane.

The fabrication process for the tested grating followed the reflection grating fabrication process described in \citet{Miles:2018aa}. The grating first was written onto a silicon substrate via e-beam lithography and then coated with $\sim5$ nm of Cr and $\sim15$ nm of Au via sputter deposition to increase its reflectivity at the tested wavelength. 

% An image of the resulting groove profile is presented in Figure \ref{fig:groove_profile}.

% \begin{figure}
% \centering
% \includegraphics[width=0.7\linewidth]{Scan Jun 11, 2020 at 10.29 AM.pdf}
% \caption{Artist's representation of an atomic force microscopy (AFM) surface measurement of the tested grating. }
% \label{fig:groove_profile}
% \end{figure}

\begin{figure*}
\centering
\includegraphics[width=\linewidth]{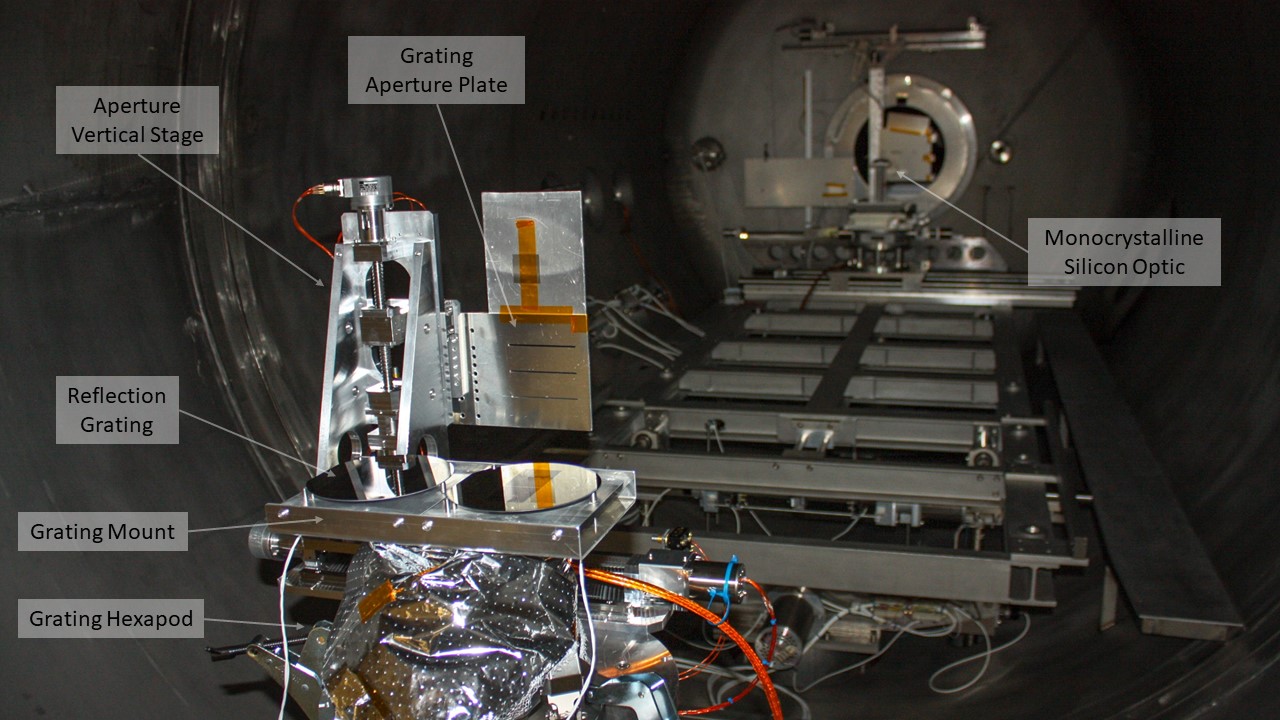}
\caption{The grating under test and grating aperture plate on top of a pedestal (not visible) in the PANTER test chamber. The mono-crystalline silicon optic used in the test campaign can be seen in the background of the image.}
\label{fig:grating_image}
\end{figure*}

The grating was mounted onto a 6-axis hexapod in the PANTER test chamber. Between the optic and the grating was an aperture plate with a 1.31 mm tall aperture. This aperture was used to reduce the size of the converging optic beam at the grating. This converging beam had predicted dimensions from raytrace modeling of $\sim30$ mm wide by $\sim2.9$ mm tall, but the grating at the nominal grazing incidence angle ($\eta$) of 1.5 degrees only presented a height of 2.62 mm. Therefore, the 1.31 mm tall aperture reduced the converging beam height so that only 50\% of the patterned grating surface was illuminated. This percentage was chosen to ensure illumination of only the grating pattern and not any unpatterned portions of the substrate. The aperture plate was mounted on vertical and horizontal translation stages. Both the grating and grating aperture staging were then mounted onto a pedestal on top of a large travel horizontal stage which allowed motion into/out of the converging optic beam. The grating, hexapod, and grating aperture plate can be seen in Figure \ref{fig:grating_image}.

\subsection{Apparatus Alignment} 
\label{sect:grat_alignment}
Previous grating performance tests showed that alignment of all components in the test apparatus is imperative to place the grating in the desired test orientation, to understand the systematic uncertainties inherent to the test apparatus, for efficient testing at dispersion, and to achieve maximum grating performance \citep{Donovan:2018aa}. Therefore, the orientation of the spectrometer components relative to one another needed to be characterized and then these components needed to be aligned together. 

% The mono-crystalline silicon X-ray optic at the PANTER X-ray Test Facility had already been aligned and extensively characterized prior to the test campaign, so the following sections will not cover its alignment and characterization. 

\subsubsection{Initial Alignment}
Before pumping down the facility, an initial alignment of the grating occurred in the optical bandpass\footnote{The mono-crystalline silicon X-ray optic at the PANTER X-ray Test Facility had already been aligned and extensively characterized prior to the test campaign.}. A laser at the source end of the facility was directed down the beamline toward the test chamber. After reflecting off of the primary (paraboloid) and secondary (hyperboloid) mirrors of the X-ray optic, the laser propagated toward the grating pedestal. The grating was inserted into this converging laser beam and positioned such that the beam was centered both horizontally and vertically on the projected grating surface. The position of the zero-order LSF on the focal plane was noted and then the grazing incidence angle of the grating was decreased until the zero-order LSF overlapped the optic focus ($\eta=0$ deg). The grating was then pitched to the nominal $\eta=1.5$ deg. and stage positions corresponding to this orientation were noted. The stage positions which placed the grating aperture plate into the converging laser beam were noted as well. Finally, the distance between the grating and the focal plane was measured with a laser distance meter ($L=3248.6 \pm 1.5$ mm). This completed the initial alignment of the test apparatus and the facility was pumped down. 

\subsubsection{Optic/Detector Coordinate System Characterization}

Once the facility was pumped down, a more thorough and precise alignment of the test apparatus could be completed in the X-ray bandpass. To begin, the orientation of the detector stage axes relative to the detector coordinate system was determined. Understanding the orientation of the detector and how the detector moved on the focal plane would be crucial for all subsequent alignment and data acquisition. A sequence of frames was first taken while scanning the optic focus both vertically and horizontally across the detector with vertical and horizontal detector stage movements. By moving the detector around a stationary optic focus, the alignment of the detector stage axes could be determined relative to the detector array. Assuming that the detector horizontal and vertical pixel axes were exactly perpendicular, this characterization found that the vertical detector stage was misaligned to the detector y-axis by $+0.6\pm1.9$ degrees ($2\sigma$ error) and the horizontal detector stage was misaligned to the detector x-axis by $+0.56\pm0.03$ degrees ($2\sigma$ error). Thus, the horizontal and vertical detector stages were consistent with perpendicular ($90.0\pm1.9$ degrees; $2\sigma$ error). 

%The horizontal and vertical detector stages were taken to be perpendicular for all subsequent alignment and testing.

The alignment of the optic coordinate system and the detector on the focal plane also needed to be characterized, as the grating had to be precisely aligned to the optic to achieve maximal performance. This characterization was accomplished by observing the ``shadow'' of the X-ray optic produced by the fraction of photons from the source not interacting with the optic but still passing through the optic aperture (producing a shadow of the optic on the focal plane). Using this shadow, the azimuthal midpoint of the optic projected onto the focal plane was determined. This position, in conjunction with the optic focus position on the focal plane, gave the optic's plane of reflection relative to the detector coordinate system. This plane of reflection was determined to be misaligned relative to the detector array by $0.6\pm1.9$ degrees ($2\sigma$ error), with errors dominated by the uncertainty on the vertical stage's orientation. With the orientation of the optic known relative to the detector array and its staging, the grating and its aperture mask could then be aligned relative to the optic. 

\subsubsection{Aperture Mask Alignment}

The 1.31 mm tall grating aperture reduced the vertical extent of the converging optic beam to $\sim1.31$ mm at the grating. This reduction in size ensured that the converging optic beam only sampled the patterned portion of the grating substrate (vertical extent of the 100 mm long tested grating at $\eta=1.5$ deg. is 2.62 mm). A misalignment of this aperture would allow a portion of the converging optic beam to sample the unpatterned surface of the grating substrate, which would thwart the comparison of the observed zero-order reflection and the diffracted orders during data analysis (to be further described in Section \ref{sect:data_red_anal}). The grating aperture was aligned vertically to the converging beam by noting the X-ray flux seen at the optic focus as a function of aperture height. If the aperture plate completely blocked the converging beam, no photons would be seen on the detector. However, as the aperture plate moved vertically and the aperture entered the converging beam, the flux on the detector would increase to a maximum value. Continuing to move the aperture vertically would then again decrease the flux seen on the detector as the aperture moved out of the converging beam again. This behavior was observed for the 1.31 mm tall aperture on the aperture plate, and the vertical position of maximum zero-order flux was adopted as the nominal grating aperture vertical position. Horizontal alignment of the aperture mask was completed during initial alignment using the laser beam reflecting off of the X-ray optic.

\subsubsection{Grating Alignment}

With the optic, detector, detector stages, and grating aperture aligned, the final spectrometer component that needed to be aligned was the grating, which needed to be aligned relative to the optic in all six degrees of freedom. Misalignments in roll (rotation about $\hat{z}$ in Figure \ref{fig:grating_geometry1}a) and $\hat{z}$ would impact the achievable resolution, while misalignments in pitch (rotation about $\hat{x}$) and yaw (rotation about $\hat{y}$) would grossly affect the position of the orders on the focal plane, moving the test configuration away from the OGRE diffraction geometry. Alignment in $\hat{x}$ and $\hat{y}$ places the grating optimally in the converging optic beam to ensure proper illumination.

The grating was first aligned in its translational degrees of freedom. To begin, the grating was moved into the location which corresponded to its nominal diffraction geometry as determined during initial alignment with the laser. The grating was then moved vertically, noting the flux in zero-order as seen on the focal plane similar to the procedure used for grating aperture alignment. With this flux scan, the nominal center height of the grating substrate corresponded to where flux in zero-order was maximal. A laser distance meter measurement during initial alignment placed the grating in a location consistent with $L=3250$ mm (measured distance; $L=3248.5 \pm 1.5$ mm), so this position was not adjusted. Finally, $\hat{x}$ alignment was also completed during initial laser alignment and was not adjusted in the X-ray bandpass. 

Next, the pitch and roll of the grating were aligned relative to the converging beam of the telescope. To align the grating in pitch, its pitch was first reduced from the nominal grazing incidence angle of 1.5 degrees until the zero-order reflection and the optic focus fell on the same detector ($\eta<0.15$ deg.). The grazing incidence angle was further reduced and a plate scale for zero-order LSF position on the detector as a function of pitch angle was determined. The grating was then moved to the nominal incidence angle of zero degrees as determined by the previously calculated plate scale. Centroid estimates place zero-order and optic focus within 5 pixels of one another, which results in an estimated pitch misalignment of $\lessapprox20$ arcsec. The zero-order LSF was then maneuvered into its nominal grazing incidence angle of 1.5 degrees. Once aligned in pitch, the grating was manipulated in roll such that the zero-order LSF fell onto the imaginary radial line connecting the optic azimuthal midpoint and the optic focus centroid. This ensured that the projected grating plane on the focal plane was perpendicular to the reflection direction of the optic, which is the nominal orientation of the grating relative to the optic in roll. Estimates from the final zero-order LSF centroid indicate that the roll of the grating was aligned to within $\lessapprox1.5$ arcmin. 

Finally, the centroids of the diffracted orders were used to align the grating in yaw. At a yaw of zero degrees relative to the converging optic beam, opposing diffracted orders ($\pm$) are at the same position in the cross-dispersion direction ($\hat{y}$ in Figure \ref{fig:grating_geometry1}a). This direction is perpendicular to the radial direction connecting the optic focus and the optic azimuthal midpoint on the focal plane. Positions of the $n=\pm1$ orders were first measured and then the grating was manipulated in yaw to place these orders at roughly the same position in the cross-dispersion direction. This process was then repeated for the $n=\pm2$ and $n=\pm3$ orders. The $n=\pm3$ orders were placed to within $\lessapprox7$ detector pixels in the cross-dispersion direction which constrained yaw alignment of the grating to within $\lessapprox20$ arcsec. This completed the yaw alignment for the grating, and the grating was now aligned relative to the optic and detector.

\subsection{Test Plan} \label{sect:test_plan}
\begin{figure*}
\centering
\includegraphics[width=0.5\linewidth]{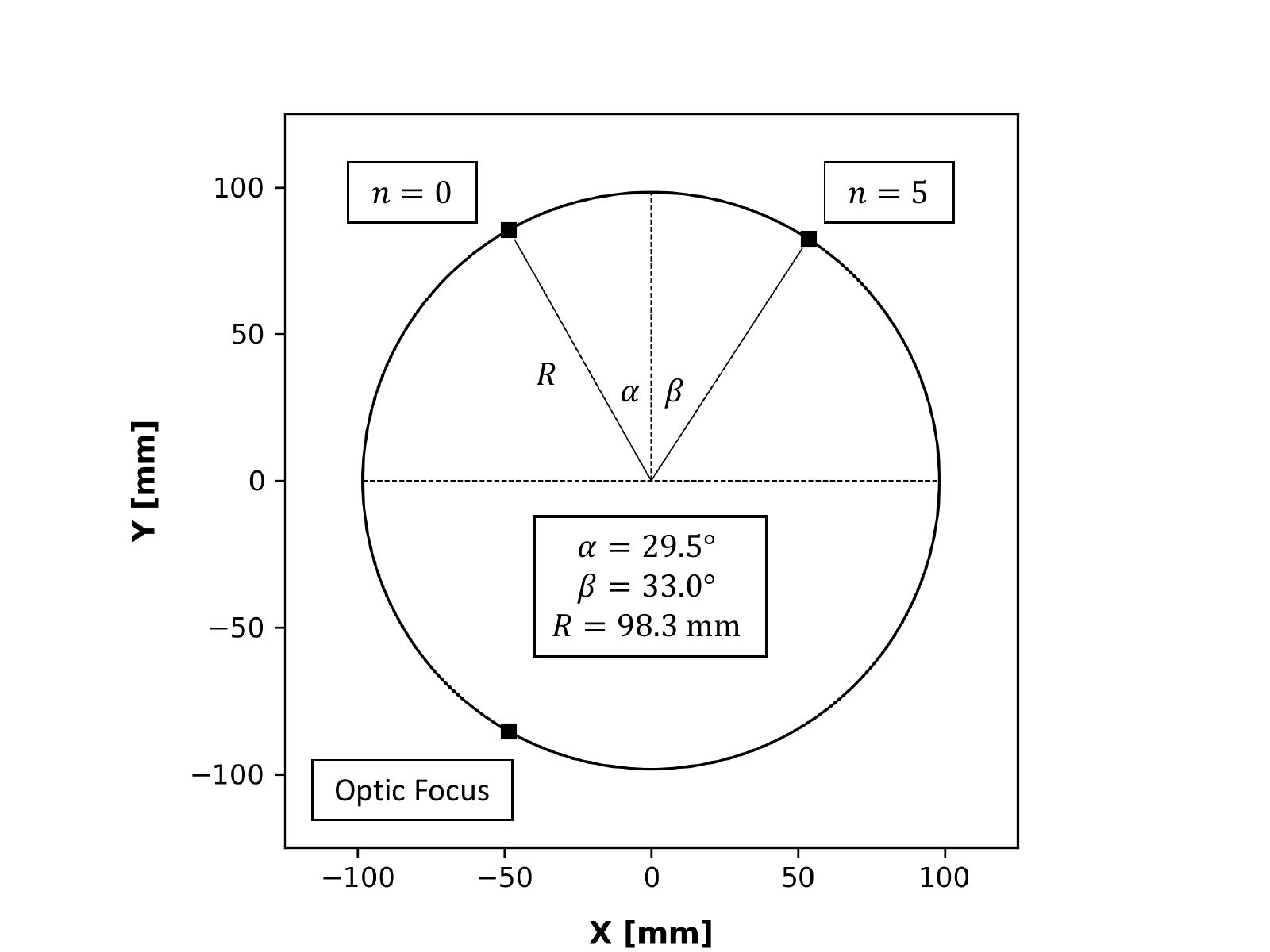}
\caption{The ideal diffraction geometry of the spectrometer under test on the focal plane. Depicted on the arc is the focus formed by the X-ray optic, the zero-order ($n=0$) reflection off of the grating surface, and the diffracted-order line-spread function ($n=5$). Also shown are the following diffraction parameters: the radius of the diffraction arc ($R$), the azimuthal angle of the incident rays ($\alpha$), and the azimuthal angle of the diffracted rays ($\beta$).}
\label{fig:diff_geom}
\end{figure*}

The spectrometer was tested in a configuration similar to the OGRE grating geometry \citep{Donovan:2019aa}. The OGRE spectrometer maximizes diffraction efficiency at $n\lambda_{b}\approx4.77$ nm; however, the available source anodes at the PANTER X-ray Test Facility precluded testing at this wavelength. The gratings could achieve a similar diffraction geometry though using fifth-order ($n=5$) Mg-K$\alpha$ ($n\lambda\sim4.95$ nm) which diffracts to $<3$ mm from the nominal OGRE blaze wavelength position on the focal plane. The electron impact source, therefore, utilized a Mg anode for this test campaign to produce Mg-K$\rm{\alpha}$ photons. The Mg-K$\rm{\alpha}$ line complex is composed of two primary components, Mg-K$\rm{\alpha_{1}}$ and Mg-K$\rm{\alpha_{2}}$, with wavelengths of $0.9889573\pm0.0000087$ nm and $0.9891553\pm0.0000103$ nm respectively in a 2:1 intensity ratio \citep{Schweppe:1994}. Table \ref{tab:diff_geom} describes the parameters for the tested diffraction geometry. A visualization of the diffraction geometry resulting from the parameters in Table \ref{tab:diff_geom} on an idealized focal plane is presented in Figure \ref{fig:diff_geom}. 

\begin{wstable}[ht]
\caption{Parameters describing the nominal diffraction geometry of the spectrometer under test.}
\begin{tabular}{@{}cccc@{}} \toprule
Parameter & Symbol & Value & Unit \\
\colrule
Graze & $\eta$ & $1.50$ & deg. \\
Roll & -- & $0$ & deg. \\
Yaw & $\Psi$ & $0.85$ & deg. \\
Cone Angle & $\gamma$ & $1.73$ & deg. \\
Throw & $L$ & $3251.49$ & mm \\
Groove Period & $d$ &  $157.57$ & nm \\
Wavelength & $\lambda$ & $\sim0.989$ & nm \\
Order & $n$ & 5 & -- \\
Dispersion & $d\lambda/dx$ & $\sim0.0097$ & nm/mm \\ \botrule
\end{tabular}
\label{tab:diff_geom}
\end{wstable}

Following alignment activities, the spectrometer was placed into the geometry described in Table \ref{tab:diff_geom} and integrations were taken of the optic PSF, the zero-order LSF, and the fifth-order LSF. Prior to each of these integrations, a focus scan was completed along the optical axis to find the position of best focus.

\section{Data Reduction \& Analysis}
\label{sect:data_red_anal}

\subsection{Data Reduction}
\begin{figure*}
    \centering
    \begin{subfigure}[ht]{0.9\linewidth}
        \includegraphics[width=\linewidth]{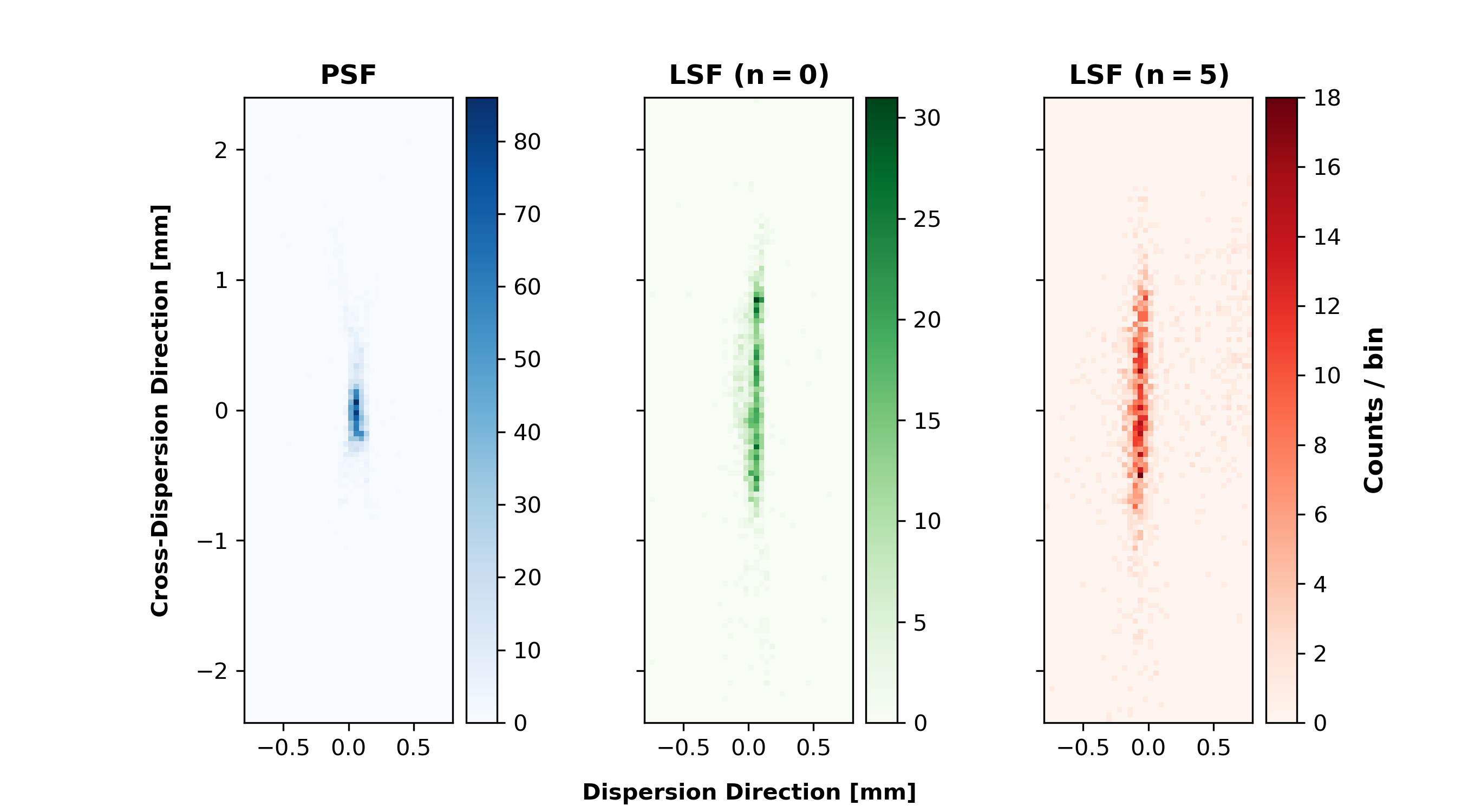}
        \caption{}
    \end{subfigure} \\[1ex]
    \begin{subfigure}[ht]{0.6\linewidth}
        \includegraphics[width=\linewidth]{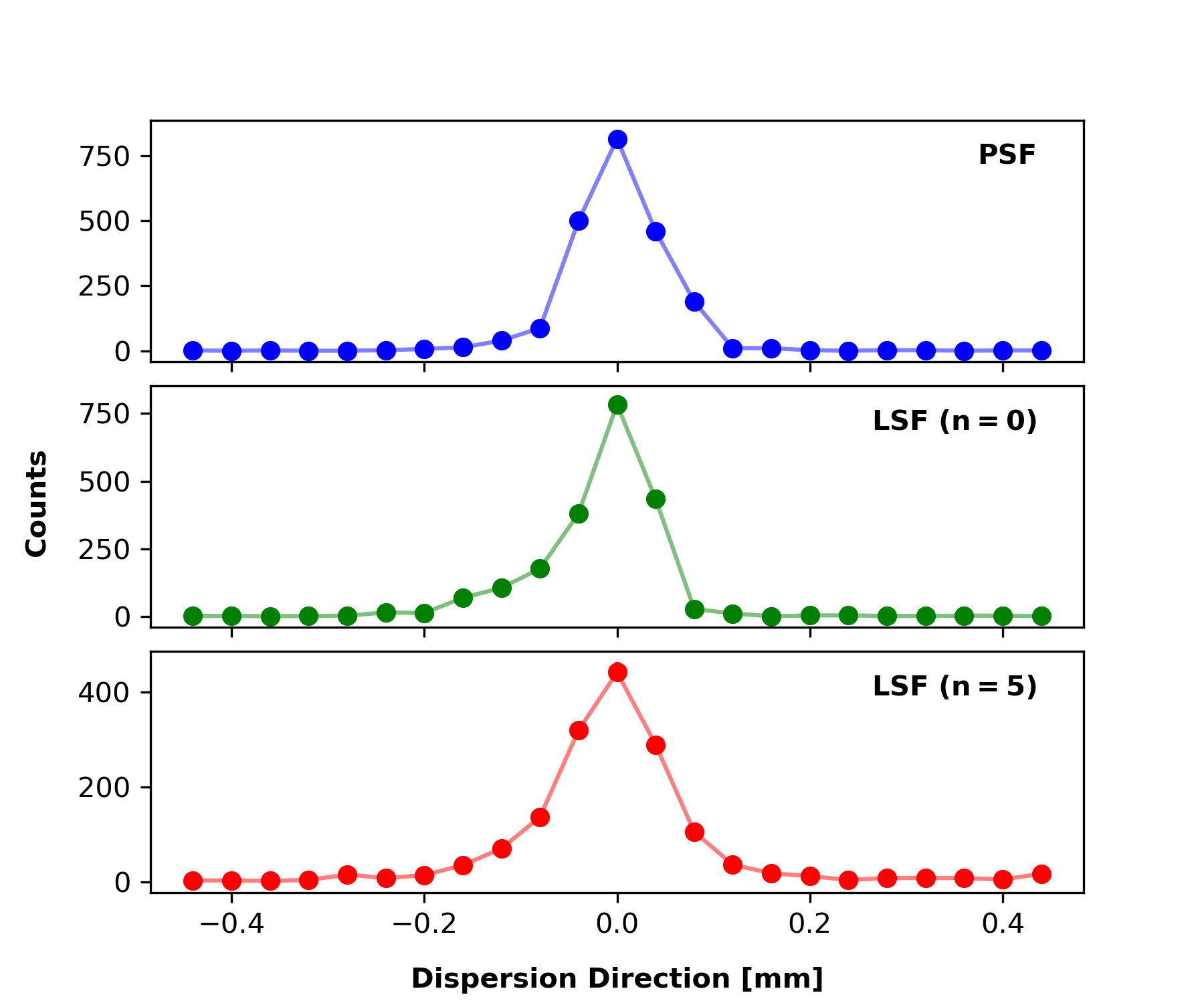}
        \caption{}
    \end{subfigure}
\caption{Data for the optic PSF (blue), zero-order LSF (green), and fifth-order LSF (red). (a) Integrated detector frames cropped around an area of interest. (b) Integrated detector frames collapsed onto the dispersion direction axis. These profiles adhere to Poisson statistics, where the error on a bin with $N$ counts is $\pm\sqrt{N}$.}
\label{fig:grat_lsf_profile}
\end{figure*}

TRoPIC reads out at a frame rate of $20$ Hz, so each integration obtained in this test campaign consisted of many individual detector frames. This technique allowed the detector to capture photon events in a frame without multiple events piling up within a given pixel. Once the detector frames for a given integration were obtained, the TRoPIC data reduction pipeline was applied to each to identify the location of each event \citep{Dennerl:2012aa}. This pipeline examines the charge distribution of individual events on the detector to identify the centroid of each photon to a subpixel resolution of 40 \si{\micro}m. Once all photons were identified for a given integration, they were then combined to form the integrated detector frames presented in Figure \ref{fig:grat_lsf_profile}a. 

% \begin{figure*}
%     \centering
%     \begin{subfigure}[ht]{0.7\linewidth}
%         \includegraphics[width=\linewidth]{pant-raw-images.pdf}
%         \caption{}
%     \end{subfigure} \\[1ex]
%     \begin{subfigure}[ht]{0.7\linewidth}
%         \includegraphics[width=\linewidth]{wald-raw-images.pdf}
%         \caption{}
%     \end{subfigure}
% \setcounter{figure}{4}   
% \caption{The observed optic focus PSF, zero-order LSF, and fifth-order LSF for Grating \#1 (a; direct write) and Grating \#2 (b; distorted write). Different zero-order and fifth-order LSF vertical extents for Grating \#1 and Grating \#2 are attributed to scatter induced by groove facet surface roughness and grating figure-induced aberrations.}
% \label{fig:raw-data}
% \end{figure*}

To form the LSF profile of each integration in the dispersion direction, each integration was collapsed onto the x-axis of the detector. As described in Section \ref{sect:grat_alignment}, this axis was found to be slightly misaligned with the nominal dispersion direction ($\Delta\theta=0.6\pm1.9$ degrees; $2\sigma$ level). However, the large uncertainties precluded the precise determination of the dispersion direction relative to the detector x-axis and the two axes were consistent at the $2\sigma$ level. Therefore, the x-axis of the detector was adopted as the dispersion direction for the grating. The resulting collapsed optic PSF, zero-order LSF, and fifth-order LSF data onto the dispersion axis are presented in Figure \ref{fig:grat_lsf_profile}b. These collapsed profiles adhere to Poisson statistics, where the error on a bin with $N$ counts is $\pm\sqrt{N}$.

\subsection{Optic Focus PSF \& Zero-Order LSF Analysis}
\label{sect:optic_zero_comp}

% The spectral resolution of a reflection grating spectrometer is limited by several factors, including the performance of the optic that is feeding the system, the size of the X-ray generating source, slope errors that are introduced by the figure of the reflection grating substrate, and diffraction-related errors. These all work to increase the extent of the diffracted-order LSF in the dispersion direction. Contributions to the diffracted-order LSF introduced by the X-ray optic and the size of the X-ray source can be seen in the optic PSF, whereas figure errors introduced by the grating substrate appear in the zero-order LSF -- a convolution of the optic PSF and errors introduced by the reflection of the optic PSF off of the grating substrate. Therefore, 

The optic PSF and zero-order LSF profiles were first analyzed to determine the impact of optic performance, the size of the X-ray source, and grating figure errors on the achievable spectral resolution of the assembled spectrometer. Contributions from diffraction-induced errors ($R_g$) will be discussed in Section \ref{sect:lsf_model}.

% Additionally, the performance of a spectrometer can be affected by the amount of data collected. However, this does not affect the inherent performance of the spectrometer, but the uncertainty on that performance. Section \ref{sect:data_red_anal} pertains only to the data collected in Figure \ref{fig:grat_lsf_profile}, and does not address the impact of statistics on spectrometer performance. This is addressed in Section \ref{sect:discussion}.

The resulting optic PSF obtained in the PANTER X-ray Test Facility and its summed profile in the dispersion direction can be seen in Figure \ref{fig:grat_lsf_profile}. As is typical of subapertured X-ray optics, the resulting optic PSF was much wider in the reflection plane (cross-dispersion direction) than in the perpendicular plane (dispersion direction). The spectral information is contained in the dispersion direction, so the width of the optic PSF in this dimension is the first error introduced into the spectrometer that limits its achievable resolution. The dispersion profile of the tested optic in Figure \ref{fig:grat_lsf_profile} was measured to be $94\substack{+8 \\ -9}$ \si{\micro\metre} FWHM. At the tested dispersion of $\sim{102}$ mm, this PSF limits the spectral resolution of the system to $R(\lambda/\Delta\lambda=x/\Delta x)\approx1080$. It is important to note that the resulting optic PSF obtained from this test campaign is a convolution of the inherent performance of the optic and the size of the X-ray source. The size of the X-ray source was not measured in this test campaign, so the relative contributions of the optic and the X-ray source to the measured optic PSF are unknown.

%\sout{The optic tested in this test campaign was an early prototype of the mono-crystalline X-ray optic technology. These optics now achieve angular resolutions of $<1.3$ arcsec HPD \cite{Zhang:2019aa} and have demonstrated widths of $<13.5$ \si{\micro}m FWHM in the dispersion direction. If the system was tested instead with more recent examples of the mono-crystalline silicon X-ray optic technology, the system resolution limit due to the finite width of the X-ray optic would have been $R>7500$.}

\begin{figure*}
\centering
\includegraphics[width=0.5\linewidth]{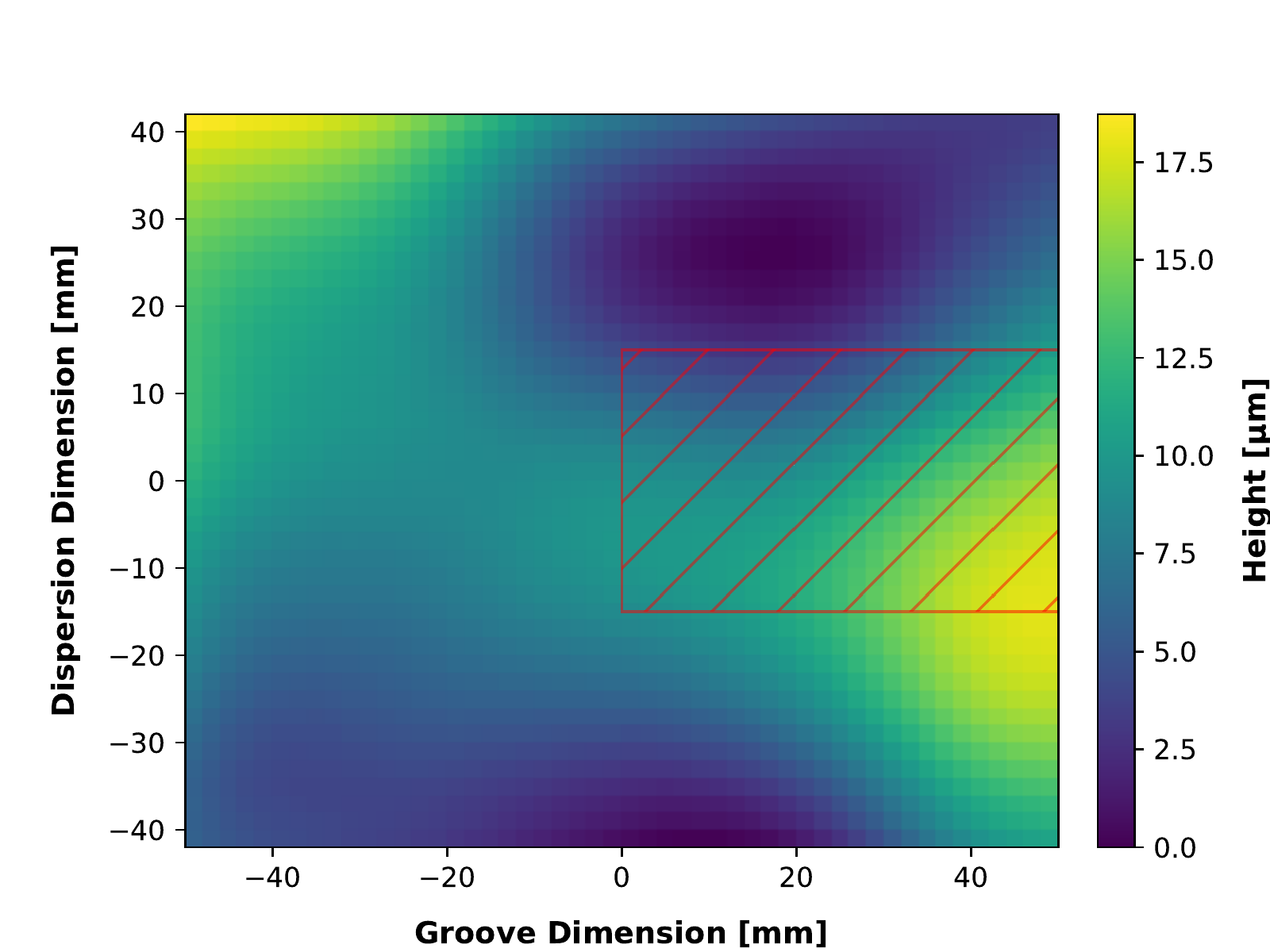}
\caption{The figure of the OGRE reflection grating prototype as measured by an optical profilometer at the Max Planck Institute for Extraterrestrial Physics. Measurements were taken with the profilometer on a $2$\:mm x $2$\:mm grid. The ``groove dimension'' denotes the direction of convergence for the grooves. Positive on the groove dimension axis points approximately toward the optic and negative on this same axis points towards the focal plane. The red hatched region represents the predicted landing location of the incident photons from the X-ray optic.}
\label{fig:grat_fig}
\end{figure*}

Slope errors introduced by the grating figure combine with the optic PSF to form the zero-order LSF. In general, these slope errors act to increase the width of the resulting zero-order LSF profile on the focal plane when compared to the optic PSF profile. This additional error further limits the spectral resolution of the system. The figure of the grating substrate as measured by an optical profilometer is presented in Figure \ref{fig:grat_fig}. While the zero-order LSF was expected to be wider than the optic PSF in the dispersion direction, this width was actually measured to be smaller than the optic PSF at $76\substack{+4 \\ -5}$ \si{\micro\metre} FWHM. This performance equates to a system resolution at dispersion of $R\approx1340$. The measured zero-order LSF and its summed profile in the dispersion direction can be seen in Figure \ref{fig:grat_lsf_profile}.

% Before addressing this discrepancy, it is important to address the uncertainty on the measured FWHM of both the optic PSF and the zero-order LSF. These uncertainties are dominated by the uncertainties on the dispersion profiles themselves. Each profile had only $\sim1500-2000$ total counts with individual subpixel bins having $<750$ total counts. These bins adhere to Poisson statistics; therefore, bins with $<750$ total counts have  uncertainties ($\sqrt{N}/N$) of $>3.5$\%. The uncertainties on the dispersion profiles, and therefore on the measured FWHMs, could be improved in the future by integrating for longer periods of time to obtain more data. 

It is peculiar that the zero-order LSF was measured to be narrower than the optic PSF in the dispersion direction. One possible explanation for this observed behavior is that the slope errors of the grating surface presented in Figure \ref{fig:grat_fig} were distributed such that they systematically narrowed the zero-order LSF. To explore this possibility, a raytrace simulation of the system was constructed.

\begin{figure*}
    \centering
    \begin{subfigure}[ht]{0.48\linewidth}
        \includegraphics[width=\linewidth]{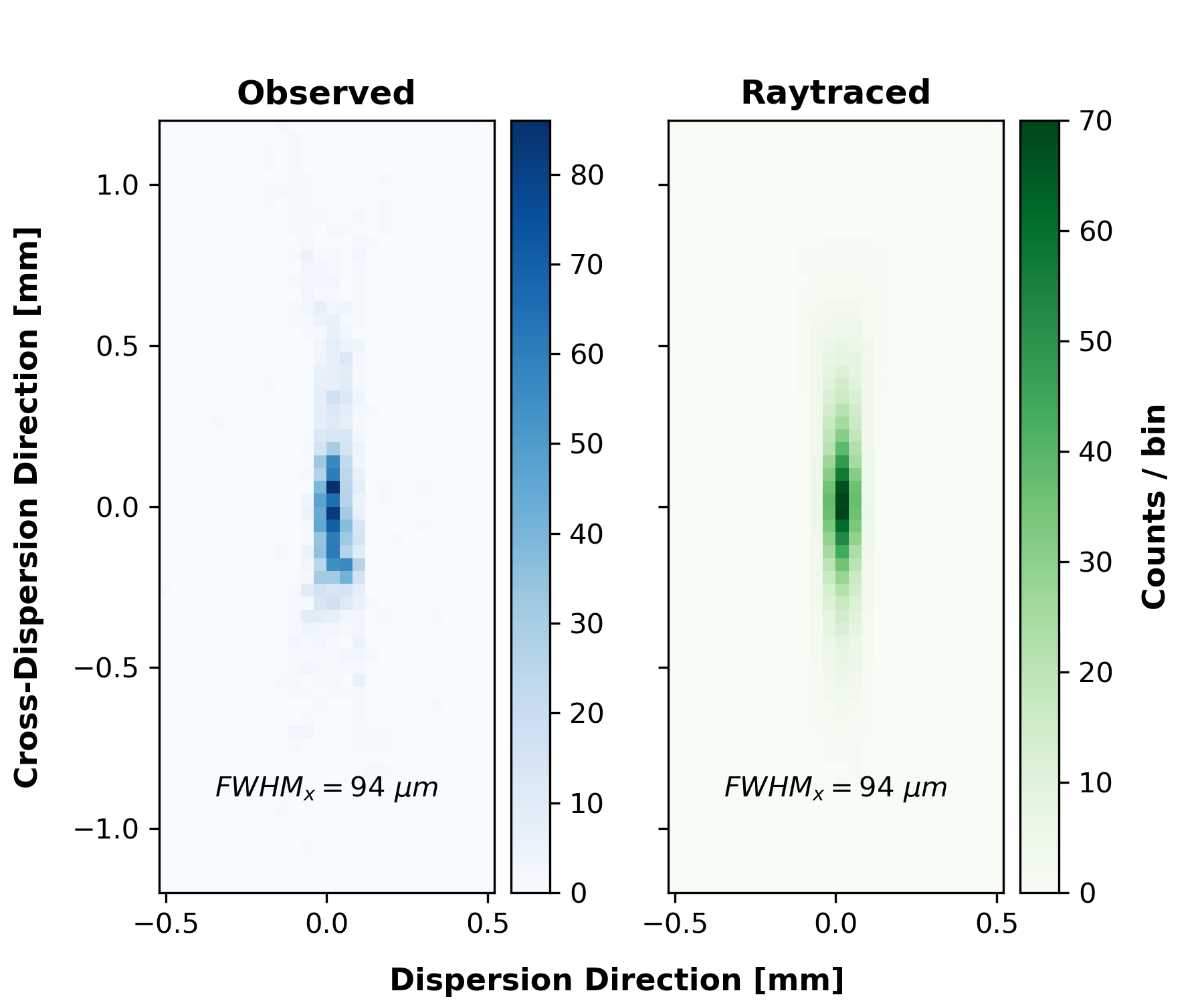}
        \caption{}
    \end{subfigure}
    \hfill
    \begin{subfigure}[ht]{0.48\linewidth}
        \includegraphics[width=\linewidth]{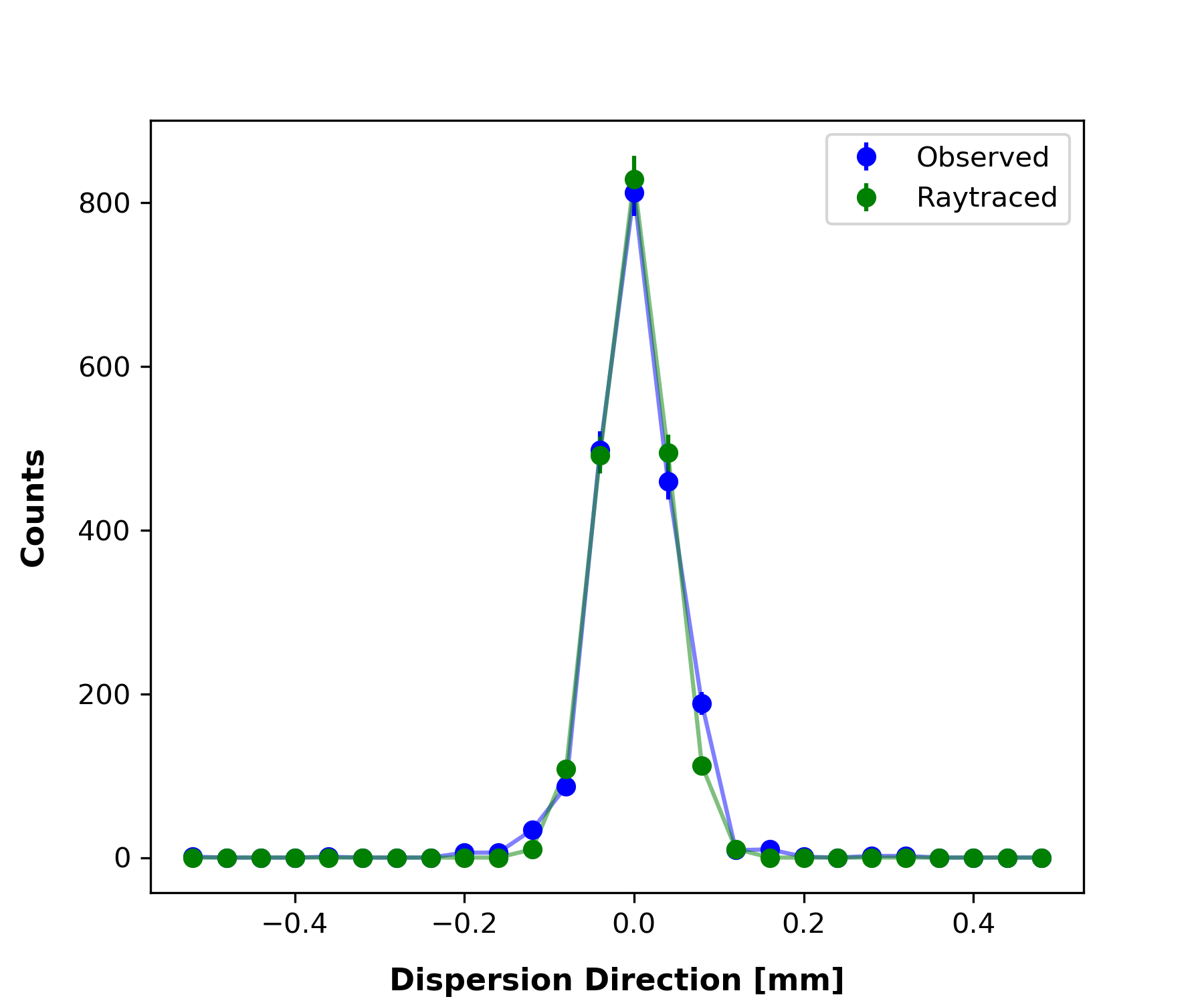}
        \caption{}
    \end{subfigure} 
\caption{A comparison of the observed (blue) and simulated (green) optic PSF (a) and the resulting PSF profile when collapsed onto the dispersion direction (b).}
\label{fig:optic_focus_comp}
\end{figure*}

An optic in accordance with the optic parameters described in Section \ref{sect:optic} was first simulated. Beckmann \citep{Beckmann:1987} and Gaussian scatter were then systematically added to the simulated rays to reproduce the shape of the observed optic PSF at the focal plane. This combined scatter function was chosen as it reproduces the characteristic ``bow-tie'' shape inherent to many experimentally tested X-ray optics, including the optic tested in this campaign \cite{Cash:1987}. As shown in Figure \ref{fig:optic_focus_comp}, the characteristic shape of the observed optic PSF in both the dispersion and cross-dispersion directions was reproduced by the raytrace. The HPDs in the cross-dispersion direction (observed: $\sim320$ \si{\micro\metre}; simulated: $\sim310$\si{\micro\metre}) and the FWHMs in the dispersion direction (observed: $94\substack{+8 \\ -9}$ \si{\micro\metre}; simulated: $94\substack{+8 \\ -7}$ \si{\micro\metre}) agree within error. Furthermore, the simulated profile in the dispersion direction has good agreement with the observed profile. 

% Before testing, the figure of the tested grating substrate was measured with an optical profilometer. These results are presented in Figure \ref{fig:grat_fig}. The red-hatched region in this figure denotes the expected landing location of the 

% The ``hump'' feature seen on the left-hand side of the zero-order LSF profile (which is not seen in the optic PSF profile) further supports this hypothesis that the figure of the grating substrate is causing the narrowing of the zero-order LSF.

% To determine if this discrepency could be explained by grating figure, the figure was first examined qualitatively. The figure map of the grating substrate as measured with an optical profilometer at the Max Planck Institute for Extraterrestrial Physics prior to installation into the test chamber can be seen in Figure \ref{fig:grat_fig}. The predicted area of illumination on the grating surface as determined via a raytrace simulation of the system is shown in this figure as a red hatched region. In the upper side of this region, photons would be interacting with the beginnings of a deep valley that could contribute to the observed hump in the zero-order LSF. Additionally, a high feature is present on the right-hand side of this region that could also contribute to this hump. 

\begin{figure*}
    \centering
    \begin{subfigure}[ht]{0.48\linewidth}
        \includegraphics[width=\linewidth]{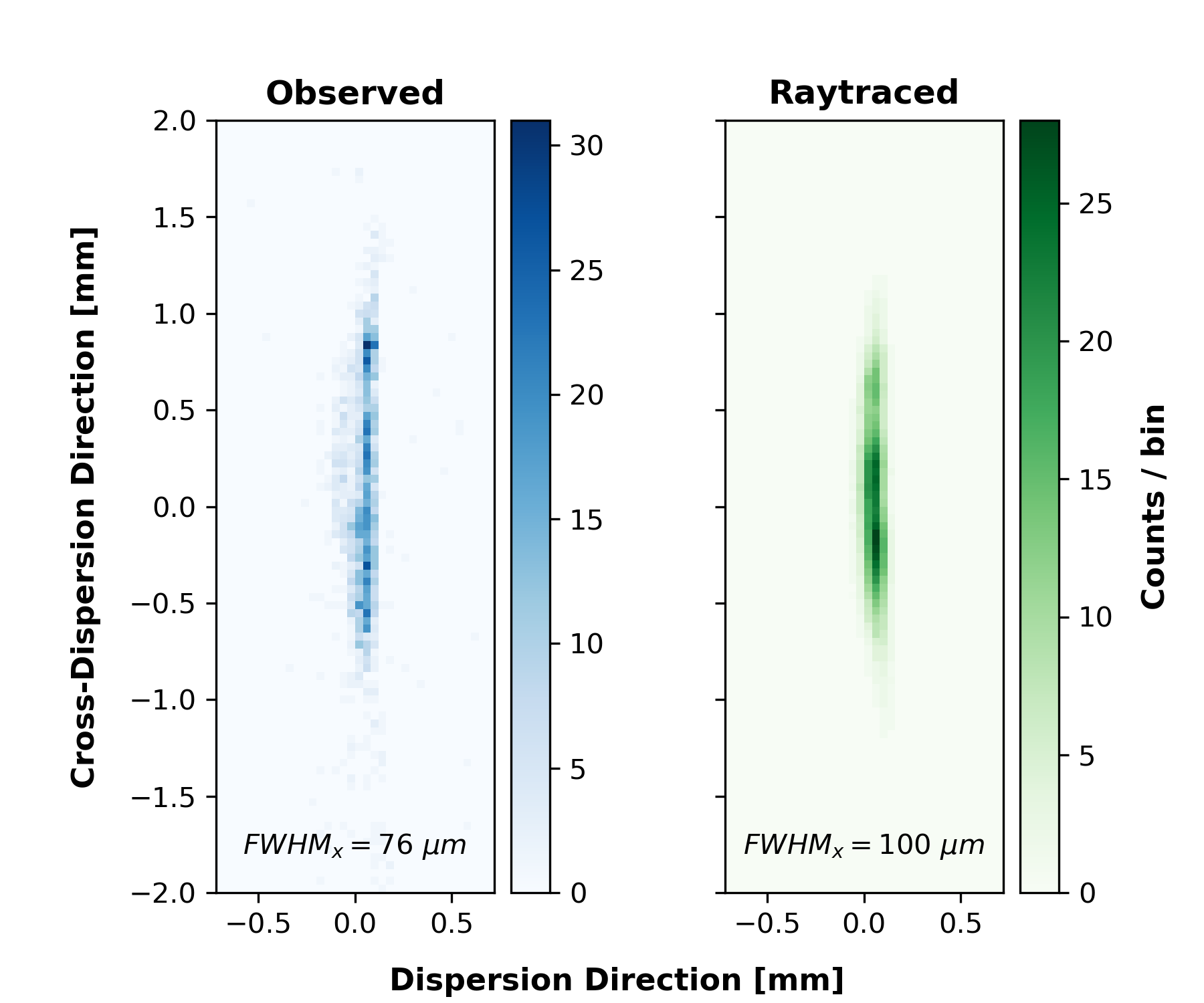}
        \caption{}
    \end{subfigure}
    \hfill
    \begin{subfigure}[ht]{0.48\linewidth}
        \includegraphics[width=\linewidth]{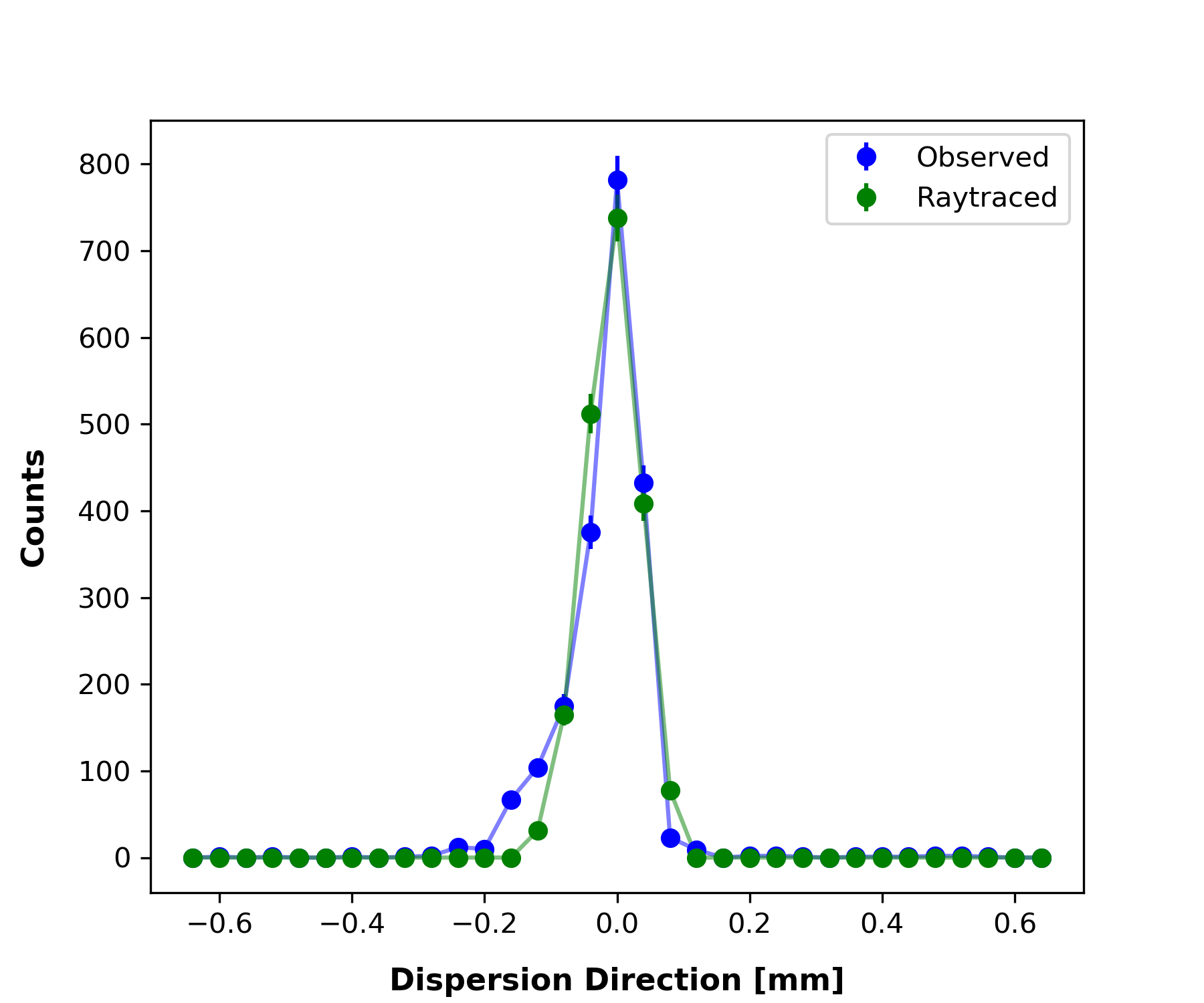}
        \caption{}
    \end{subfigure} 
\caption{A comparison of the observed (blue) and simulated (green) zero-order LSF (a) and the resulting LSF profile when collapsed onto the dispersion direction (b).}
\label{fig:zero_order_comp}
\end{figure*}

Once the optic PSF was reproduced via raytrace, a grating with figure in agreement with that in Figure \ref{fig:grat_fig} was inserted into the simulation and placed into the as-tested orientation determined during system alignment. Rays were then reflected off of this grating surface and propagated to the focal plane. The resulting simulated zero-order LSF can be seen in Figure \ref{fig:zero_order_comp} along with the observed zero-order LSF profile. The simulated profile was measured to have a width of $100\substack{+7 \\ -9}$ \si{\micro\metre} FWHM in the dispersion direction. This width does not agree with the observed zero-order LSF width of $76\substack{+4 \\ -5}$ \si{\micro\metre} FWHM within error. The full-width at quarter-maximum (FWQM) and full-width at three-quarters-maximum (FW3QM) were explored in addition to the FWHM to determine whether encapsulating more or less of the profile would result in better agreement. While the agreement was slightly better for both the FWQM (observed FWQM: $120\substack{+13 \\ -9}$ \si{\micro\metre}; simulated FWQM: $143\substack{+6 \\ -6}$ \si{\micro\metre}) and FW3QM (observed FW3QM: $48\substack{+1 \\ -4}$ \si{\micro\metre}; simulated FW3QM: $62\substack{+7 \\ -5}$ \si{\micro\metre}), the two profiles still did not agree within error. In addition to the disagreement in width, the simulation cannot reproduce the ``hump'' feature on the left-hand side of the observed profile. The simulation does, however, appear to reproduce a couple features seen in the observed zero-order LSF, specifically the slight curve in the overall shape of the zero-order LSF and the discontinuity in the upper half of the profile.

There could be several reasons why the observed and simulated zero-order LSF widths do not agree, including raytrace simulation inaccuracies, a misalignment in the grating relative to expectation, and/or a change in the grating figure. These three possibilities will be addressed in turn. First, the raytrace simulation is only an approximation of the spectrometer tested. Beckmann and Gaussian scatter were introduced into the raytrace simulation to reproduce the observed optic PSF; however, this scatter scenario is an oversimplification. While it appears to reproduce the observed optic PSF at the focal plane, it may not reproduce the ray distribution on the grating surface. Without knowing the exact scatter distribution of the rays exiting the X-ray optic, this simulation can never reproduce the observed profile exactly. Additionally, the grating could have been slightly misaligned in the cross-groove (dispersion) dimension. A misalignment in this dimension had the loosest tolerance of all grating alignment tolerances and was aligned visually prior to testing. It is estimate that this misalignment could be on the order of $\pm2-3$ mm. If the red region in Figure \ref{fig:grat_fig} were to move up $2-3$ mm, the rays would be exposed to more of the low valley shown in dark blue. This region could theoretically contribute a hump-like feature in the zero-order profile. Such a misalignment was explored via raytrace simulations, but the results were inconclusive. While the width did change given a misalignment in this dimension, it still did not agree within error with the observed width. Furthermore, a hump-like feature of the magnitude present in the observed zero-order LSF profile could never be reproduced. Finally, there could have been a change in the figure of the grating between measurement and test. The back of the grating substrate was epoxied onto three ball-pointed set screws which were screwed into a metal plate. If these set screws somehow move slightly during installation, the figure of the substrate during testing would be different than the measured figure. The raytrace simulation would therefore not be able to accurately reproduce the observed data. 

% While this explanation is the least likely of the three, it cannot be ruled out.

While the observed discrepancy between the optic PSF and the zero-order LSF is peculiar and was not expected, subsequent analysis to be described in Section \ref{sect:lsf_model}, which involves the comparison of the zero-order and fifth-order LSF profiles, is fortunately not impacted by this discrepancy. All of the postulated explanations for the discrepancy are either not physical (raytrace inaccuracies) or would be consistent across all integrations (a grating misalignment or change in grating figure after measurement), and therefore do not affect this comparison.

\subsection{Construction of a LSF Model} \label{sect:lsf_model}
The observed fifth-order LSF on the focal plane is a convolution of the optic PSF, aberrations induced by the grating substrate figure, the underlying shape of the observed spectral lines at dispersion, and aberrations introduced during diffraction. The observed grating zero-order LSF ($n=0$) represents the convolution of the optic PSF and aberrations induced by the grating figure. Therefore, the observed fifth-order LSF can be expressed as:

\begin{equation}
\label{eq:lsf_prof}
\mathrm{LSF}(n=5)=\mathrm{LSF}(n=0)*\phi_{Mg-K\alpha}*G_{diff},
\end{equation}

\noindent where $\mathrm{LSF}(n=0)$ is the observed zero-order LSF (representing the convolution of the optic PSF and aberrations induced by the grating figure), $\phi_{Mg-K\alpha}$ is the spectral line shape (in this case Mg-K\text{$\alpha_{1,2}$}), and $G_{diff}$ represents the aberrations introduced during diffraction.

The theoretical line profile of a single line in the Mg-K\text{$\alpha_{1,2}$} doublet is represented by the Lorentzian (Cauchy) profile:

\begin{equation}
L(\lambda; A, \lambda_{0}, \gamma)=\frac{A}{\pi}\left(\frac{\gamma}{(\lambda-\lambda_{0})^{2}+\gamma^{2}} \right),
\end{equation}

\noindent where $A$ is the amplitude, $\lambda_{0}$ is the center wavelength of the profile, and $\gamma$ is the half-width at half-maximum ($2\gamma\equiv\Gamma$; FWHM). The Mg-K\text{$\alpha_{1,2}$} doublet, $\phi_{Mg-K\alpha}$, can thus be expressed as a double Lorentzian profile:

\begin{equation}
\label{eq:phi}
\phi_{Mg-K\alpha}(\lambda; A, \lambda_{0}, \gamma, \Delta\lambda)=\frac{A}{\pi}\left(\frac{\gamma}{(\lambda-\lambda_{0})^{2}+\gamma^{2}}+\frac{\gamma}{2(\lambda-\lambda_{0}+\Delta\lambda)^{2}+\gamma^{2}} \right),
\end{equation}

% $\phi_{Mg-K\alpha}$ is a double-peaked Lorentzian distribution with peaks representing the summed distribution of the Mg-K\text{$\alpha_{1}$} and Mg-K\text{$\alpha_{2}$} transition lines. 

\noindent where $\Delta\lambda$ is the separation between the Mg-K\text{$\alpha_{1}$} and Mg-K\text{$\alpha_{2}$} lines. Both transition line wavelengths ($\lambda_{1}$, $\lambda_{2}$) and their widths ($\Gamma_{\lambda_{1}}$ \& $\Gamma_{\lambda_{2}}$) have been measured previously, while theory constrains their relative amplitudes to a 2:1 intensity ratio \citep{Klauber:1993, Schweppe:1994}. This 2:1 intensity ratio is implemented into Equation \ref{eq:phi}. The literature parameters adopted for $\phi_{Mg-K\alpha}$ are shown in Table \ref{tab:line_params}.

\begin{wstable}[ht]
\caption{Literature values adopted for the $\phi_{Mg-K\alpha}$ doublet profile defined in Equation \ref{eq:phi}.}
\begin{tabular}{@{}cccc@{}} \toprule
Parameter & Value (\si{\angstrom}) & Error (\si{\angstrom}) & Source \\
\colrule
$\lambda_{0}$ & 9.889573 & 0.000087 & \cite{Schweppe:1994} \\
$\Delta\lambda$ & 0.001980 & 0.000135 & \cite{Schweppe:1994} \\
$2\gamma=\Gamma$ & 0.00427 & 0.00004 & \cite{Klauber:1993} \\ \botrule
\end{tabular}
\label{tab:line_params}
\end{wstable}

% $\phi_{Mg-K\alpha}$ can expressed in physical space or in wavelength space. It doesn't much matter as he separation of Mg-K$\alpha_{1}$ and Mg-K$\alpha_{2}$ on the focal plane is given by:

% \begin{equation}
%     \Delta x=\frac{nL(\lambda_{2}-\lambda_{1})}{d},
% \end{equation}

% \noindent where $n$ is the diffraction order, $d$ is the groove period, and $L$ is the grating throw.

$G_{diff}$ represents all errors introduced during diffraction, including grating period errors and geometric aberrations. Local period errors result when the as-manufactured grating period at a given location on the grating surface does not agree with the designed period at that same location. It can be shown with Equation \ref{eq:2} that period errors cause dispersion errors, which manifest on the focal plane as a blurring of the diffracted-order LSF in the dispersion direction. Geometric aberrations can result through some combination of diffraction-related astigmatism \citep{DeRoo:2020}, measurements being taken at a non-optimal focal plane, the effects of the flat detector being used to sample a curved focal plane, and grating misalignments. These aberrations also work to broaden the diffracted-order LSF on the focal plane. In this analysis, this diffraction-induced blur is modelled as a Gaussian profile. This allows Equation \ref{eq:lsf_prof} to be rewritten as:

\begin{equation}
\label{eq:lsf_prof2}
\mathrm{LSF}(n=5)=\mathrm{LSF}(n=0)*(\phi_{Mg-K\alpha}*G_{diff}),
\end{equation}

\noindent where ($\phi_{Mg-K\alpha}*G_{diff}$) -- the convolution of the double-peaked Lorentzian profile $\phi_{Mg-K\alpha}$ and single-peaked Gaussian profile $G_{diff}$ -- is a double-peaked Voigt profile. The remaining two components in Equation \ref{eq:lsf_prof} are $\mathrm{LSF}(n=0)$ and $\mathrm{LSF}(n=5)$ -- the collapsed profiles shown in Figure \ref{fig:grat_lsf_profile}. Thus, all components in Equation \ref{eq:lsf_prof} are known except for $G_{diff}$. 

\subsection{A Bayesian Model to the Diffracted-Order LSF} \label{sect:model_fitting}
Bayesian statistics were used to construct the LSF model and to determine $G_{diff}$, as it provided a framework to include prior information on previously measured model parameters and other uncertainties present in the model. Bayesian statistics is built upon Bayes' theorem:

\begin{equation} \label{eq:bayes}
P(\theta|D)=\frac{P(D|\theta) P(\theta)}{P(D)},
\end{equation}

\noindent where $D$ represents the observed data to be analyzed and $\theta$ represents the parameters of the model. $P(D|\theta)$ then represents the probability of obtaining the data for a given set of model parameters (\textit{the likelihood}), $P(\theta)$ represents the prior information on the probability of the model parameters (\textit{the prior}), and $P(D)$ represents the probability of the data for all models (\textit{the evidence}). These three terms combine as described in Eq. \ref{eq:bayes} to give $P(\theta|D)$, the probability of obtaining the given model parameters for some given data (\textit{the posterior}). Since $P(D)$ does not depend on the model parameters, it is essentially just a scaling term. Therefore, Bayes' theorem can also be written as:

\begin{equation}
P(\theta|D)\propto P(D|\theta) P(\theta).
\end{equation}

In the context of model fitting, Bayes' theorem takes in any prior information on the model parameters $P(\theta)$ and the likelihood $P(D|\theta)$ to construct a posterior probability distribution $P(\theta|D)$. The prior information can be any information known previously from theory, observations, or measurements. In the case of the LSF model described in Equation \ref{eq:lsf_prof}, $P(\theta)$ includes prior information on the Mg-K\text{$\alpha_{1,2}$} line profile and the measured zero-order and fifth-order LSF profiles. The likelihood is akin to the likelihood function found in frequentist statistical methods such as maximum likelihood estimation where it represents the plausibility of the model given the observed data. The product of the prior and the likelihood then gives a posterior probability distribution of the model parameter space. For simple models, this can be calculated analytically or numerically, but large parameter spaces quickly reach a computational problem \citep{VanderPlas:2014aa}.

Markov chain Monte Carlo (MCMC) sampling methods solve the computation problem of Bayesian modeling by generating samples from the posterior probability distribution given a supplied prior and likelihood. With a large number of samples, the sampled posterior probability distribution will converge to the true posterior probability distribution. Numerous MCMC sampling methods exist as well as implementations of these methods in various programming languages. For this data analysis, a Python implementation of the Metropolis-Hastings sampler, PyMC \citep{Patil:2010aa}, was used to sample the posterior distribution.

\subsection{Modeling Zero-Order LSF}
Prior to fitting the fifth-order LSF profile, a model for the observed zero-order LSF profile was required. The true model for $\mathrm{LSF}(n=0)$ is believed to be made up of one or more peak components, but the exact number of components is unknown. Therefore, a grid of models were fit to the observed zero-order data and a best-fit zero-order model was chosen from this grid of models. This procedure is similar to that presented in \citet{DeRoo:2020}. Each location on the grid ($N_{L}$, $N_{G}$) represents a zero-order model with a specific number of Gaussian components ($N_{G}$) and Lorentzian components ($N_{L}$). The number of total components in the zero-order model at that particular grid location is given by $N_{tot}=N_{G}+N_{L}$ where $N_{tot}\leq4$. Each Gaussian or Lorentzian component in the model can be described by an amplitude ($A$), center position ($x_{0}$), and a width ($\sigma$). The profile at a particular location on the grid is then:

\begin{equation}
\mathrm{LSF}(x; n=0, \{\theta\})=\sum_{i}^{N_{G}}\{G(x; A_{G}^i, x_{0,G}^{i}, \sigma_{G}^i)\}+\sum_{j}^{N_{L}}\{L(x; A_{L}^j, x_{0,L}^{j}, \sigma_{L}^j)\},
\end{equation}

\noindent where $\{G(x; A_{G}^i, x_{0,G}^{i}, \sigma_{G}^i)\}$ represents the Gaussian components for that particular grid position, $\{L(x; A_{L}^j, x_{0,L}^{j}, \sigma_{L}^j)\}$ represents the Lorentzian components for that particular grid position, and $\{\theta\}$ represents the parameter set for each component in the model.

\begin{figure}
\centering
\includegraphics[width=\linewidth]{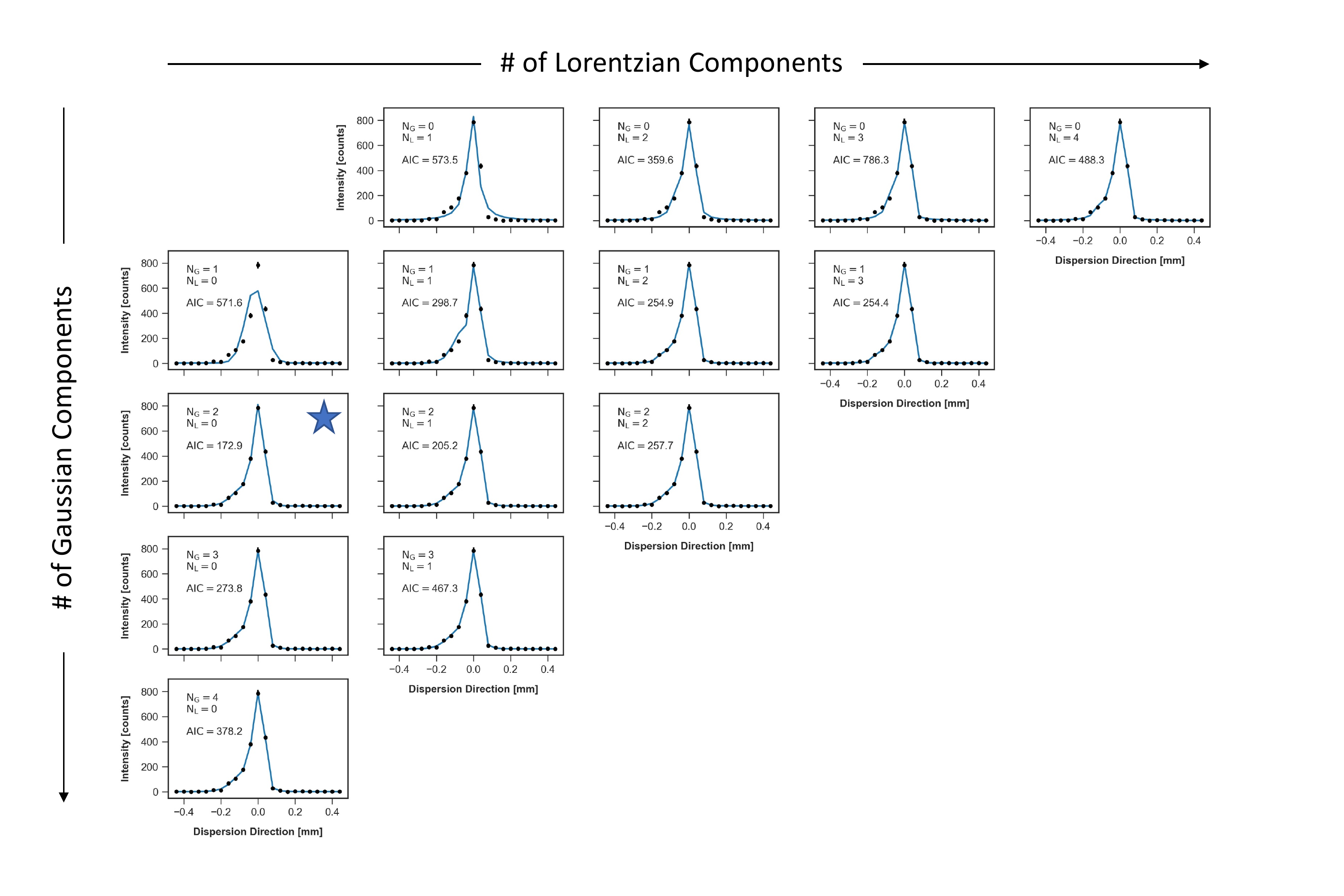}
\caption{Grid of fits to the zero-order LSF profile. Each model is composed of $N_{G}$ Gaussian components (vertical axis) and $N_{L}$ Lorentzian components (horizontal axis). For each fit, the Akaike information criterion (AIC) \citep{Akaike:1974} was calculated and is displayed at each grid location. The best fit is denoted by a blue star and is at grid location ($N_{G}=2$, $N_{L}=0$) with $\textrm{AIC}=172.9$.}
\label{fig:zero_grid}
\end{figure}

The resulting grid of zero-order LSF models is shown in Figure \ref{fig:zero_grid}. At each grid location, the Akaike information criterion (AIC) \citep{Akaike:1974} was calculated. This criterion is given by $\textrm{AIC}=2k-2\ln(\hat{L})$ , where $\hat{L}$ is the maximum value of the likelihood function and $k$ is the number of parameters in the model. The preferred model is the model with the lowest AIC value, with the AIC penalizing an overly complex model ($k$) and rewarding goodness of fit ($\hat{L}$). 

\begin{figure}
\centering
\includegraphics[width=0.5\linewidth]{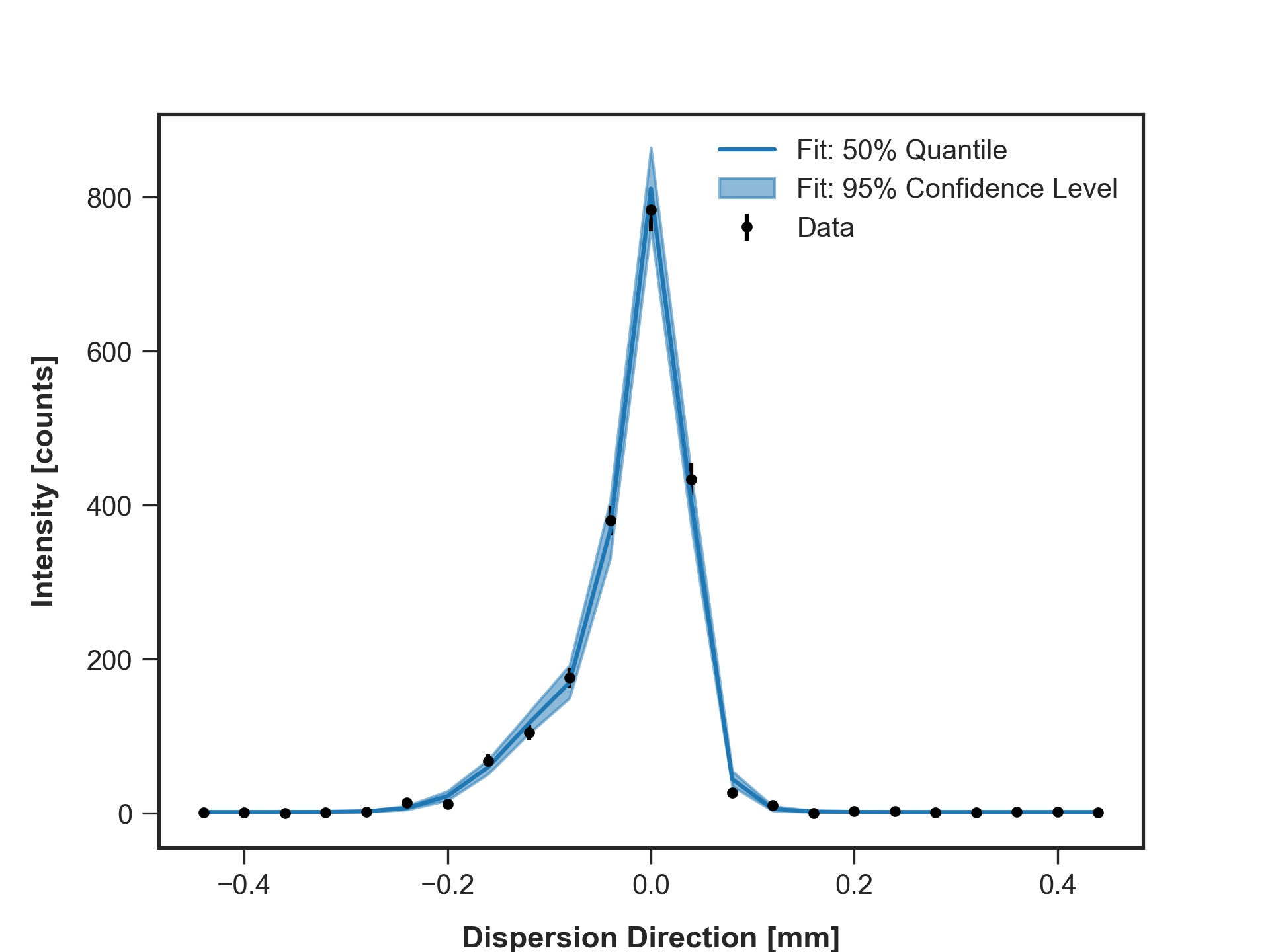}
\caption{The best-fit model to the zero-order LSF profile as determined by the Akaike information criterion (AIC) \citep{Akaike:1974}. This model is composed of no Lorentzian and two Gaussian components ($N_{L}=0$, $N_{G}=2$) and is denoted in Figure \ref{fig:zero_grid} by a blue star. }
\label{fig:zero_best_fit}
\end{figure}

The best-fit model as determined by the AIC is composed of no Lorentzian and two Gaussian components ($N_{L}=0$, $N_{G}=2$). The grid location corresponding to this model is denoted by a blue star in Figure \ref{fig:zero_grid}. This best-fit model is also shown separately in Figure \ref{fig:zero_best_fit}. With a best-fit zero-order model identified, it could then be incorporated into the fifth-order LSF model to determine $G_{diff}$. 
 
%Discussion of the suitability of this sampling method is beyond the scope of this paper.

\subsection{Diffracted-Order LSF Fitting \& Results} \label{sect:results}

The parameters which define the fifth-order LSF model are shown in Table \ref{tab:prior_table}. Also shown in this table are the prior distributions that are assigned to each parameter. These prior distributions, as outlined in Section \ref{sect:model_fitting}, describe the prior knowledge on each parameter. All measured parameters ($\lambda_{1}$, $\Delta\lambda$, $\Gamma_{1,2}$, and $d\lambda/dx$) were assumed to be represented by a Gaussian probability distribution centered about the measured value of the parameter with a standard deviation defined by the stated measurement error. These parameters therefore favored the measured value, but also allowed values that were consistent with the stated measurement error. \citet{Klauber:1993} and \citet{Schweppe:1994} provided measured values for $\lambda_{1}$, $\Delta\lambda$, and $\Gamma_{1,2}$. The other measured parameter -- $d\lambda/dx$ -- is the dispersion relation. This relation converts between wavelength space and physical space on the focal plane, and its value was obtained by measuring the distance between the observed zero-order and fifth-order LSFs on the focal plane in the dispersion direction.

\begin{wstable}[ht]
\caption{Prior distribution for each parameter used when constructing the model for each LSF observation.  \label{tab:prior_table}}
\begin{tabular}{@{}ccc@{}} \toprule
Parameter & Prior & Unit \\
\colrule
$\lambda_{1}$ & $N(\mu=9.889573, \sigma=0.000087)\tnote{a}$ & \si{\angstrom} \\
$\Delta\lambda$ & $N(\mu=0.001980, \sigma=0.000135)\tnote{a}$ & \si{\angstrom} \\
$\Gamma_{1, 2}$ & $N(\mu=0.00427, \sigma=0.00004)\tnote{b}$ & \si{\angstrom} \\
% $\Gamma_{}$ & $N(\mu=0.541, \sigma=0.005)$ & eV \\
$d\lambda/dx$ &  $N(\mu=0.097, \sigma=0.001)$ & \si{\angstrom}/mm \\
$\log(R_g)$ & $U(2, 6)$ & -- \\
$x_{0}$ & $U(-0.5, 0.5)$ & mm \\
$A$ & $U(10^{-3}, 10^{2})$ & counts \\
$\textrm{bkg}$ & $U(10^{-3}, 10^{1})$ & counts \\ \botrule
\end{tabular}
\begin{tablenotes}
\item[$*$] $N(\mu, \sigma)$ denotes a Normal (Gaussian) distribution, while $U(a, b)$ denotes a Uniform distribution. 
\item[a] From \citet{Schweppe:1994}.
\item[b] From \citet{Klauber:1993}.
\end{tablenotes}
\end{wstable}

% \begin{deluxetable}{lrr}
% \tablecaption{Prior distribution for each parameter used when constructing the model for each LSF observation. \label{tab:prior_table}}
% \tablecolumns{3}
% \tablenum{1}
% \tablewidth{0pt}
% \tablehead{
% \colhead{Parameter} &
% \colhead{Prior} &
% \colhead{Units}
% }
% \startdata
% $\lambda_{1}$ & $N(\mu=9.889573, \sigma=0.000087)$ & \si{\angstrom} \\
% $\lambda_{2}$ & $N(\mu=9.891553, \sigma=0.000103)$ & \si{\angstrom} \\
% $\Gamma_{\lambda_{1}}$ & $N(\mu=0.541, \sigma=0.005)$ & eV \\
% $\Gamma_{\lambda_{2}}$ & $N(\mu=0.541, \sigma=0.005)$ & eV \\
% $L$ &  $N(\mu=3248.6, \sigma=1.5)$ & mm \\
% $\sigma$ & $U(10^{-8}, 10^{2})$ & pixels \\
% $x_{0}$ & $U(30.5, 31.5)$ & pixels \\
% $A$ & $U(0, 1)$ & counts \\
% \enddata
% \tablecomments{Values for $\lambda_{1,2}$ and $\Gamma_{1,2}$ come from \citet{Schweppe:1994} and \citet{Klauber:1993} respectively. $L$ was measured using a laser distance meter during initial alignment. Prior distributions for $\sigma$, $x_{0}$, and $A$ were unknown and therefore were assigned Uniform distributions.}
% \end{deluxetable}

The remaining parameters which defined the model were the amplitude ($A$) and center ($x_{0}$) in pixel space of the double Lorentzian line profile representing the Mg-K$\alpha$ doublet line profile, the width of the Gaussian profile ($\sigma$; parameterized by $R_g=x/2.355\sigma$ in the model) representing any diffraction-induced errors, and a background term (bkg).  Each of these parameters had no prior information associated with it, so each was given a so called ``uninformative prior" which favors each value in the parameter space equally. The parameter spaces for each were chosen based upon the observed data. For example, the lower limit of $\log(R_g)$ was chosen to be $\log(R_g)=2$. This value is inconsistent with the observed data, as the diffracted-order LSF would need a width of $\sim0.4$ mm FWHM to be consistent with this value of $\log(R_g)$. The upper limit of $\log(R_g)$ was chosen to be $\log(R_g)=6$ which at the tested dispersion ($x\sim102$ mm) is consistent with $R_g=\infty$. Therefore, the $\log(R_g)$ prior of $U(2,6)$ fully encompasses all possible values of the parameter. Similar exercises were performed to assign realistic parameter spaces to bkg, $x_0$, and $A$. These parameters and their prior distributions are also presented in Table \ref{tab:prior_table}.

Once the fifth-order LSF model was constructed, samples of the posterior probability distribution were obtained via PyMC. In PyMC, three parameters are used to describe the sampled posterior probability distribution: the total number of steps in the sample chain ($N_{step}$), the number of samples thrown away in the beginning of the sample chain (burn-in period; $N_{burn}$), and the thinning parameter used on the chain ($N_{thin}$). The burn-in period allows starting values that may not be representative of the true posterior probability distribution to be discarded from the final sample, while thinning can be used to form an independent sample from the posterior if the neighboring samples in the original chain are not independent. To estimate these parameters for the constructed fifth-order LSF model, the Raftery-Lewis Diagnostic was used \citep{Raferty:1995aa}. This diagnostic takes in a sample of the posterior probability distribution and then recommends parameters to estimate a quantile, $q$, to a desired accuracy of $\pm a$ with a probability of $p$. An initial $N_{step}=10,000$ step sample of the posterior probability distribution was obtained without any thinning ($N_{thin}=0$) or a burn-in period ($N_{burn}=0$) and was passed to the Raftery-Lewis Diagnostic function in PyMC. To estimate the $q=0.025$ quantile of $\log(R_g)$ to an accuracy of $a=\pm0.01$ with a $p=0.95$ probability, the Raftery-Lewis Diagnostic recommended that $N_{step}=2,010,000$ total samples be obtained with the first $N_{burn}=10,000$ samples discarded and a thinning factor of $N_{thin}=40$ be used to produce independent an independent sample chain.

\begin{figure*}
\centering
\includegraphics[width=0.5\linewidth]{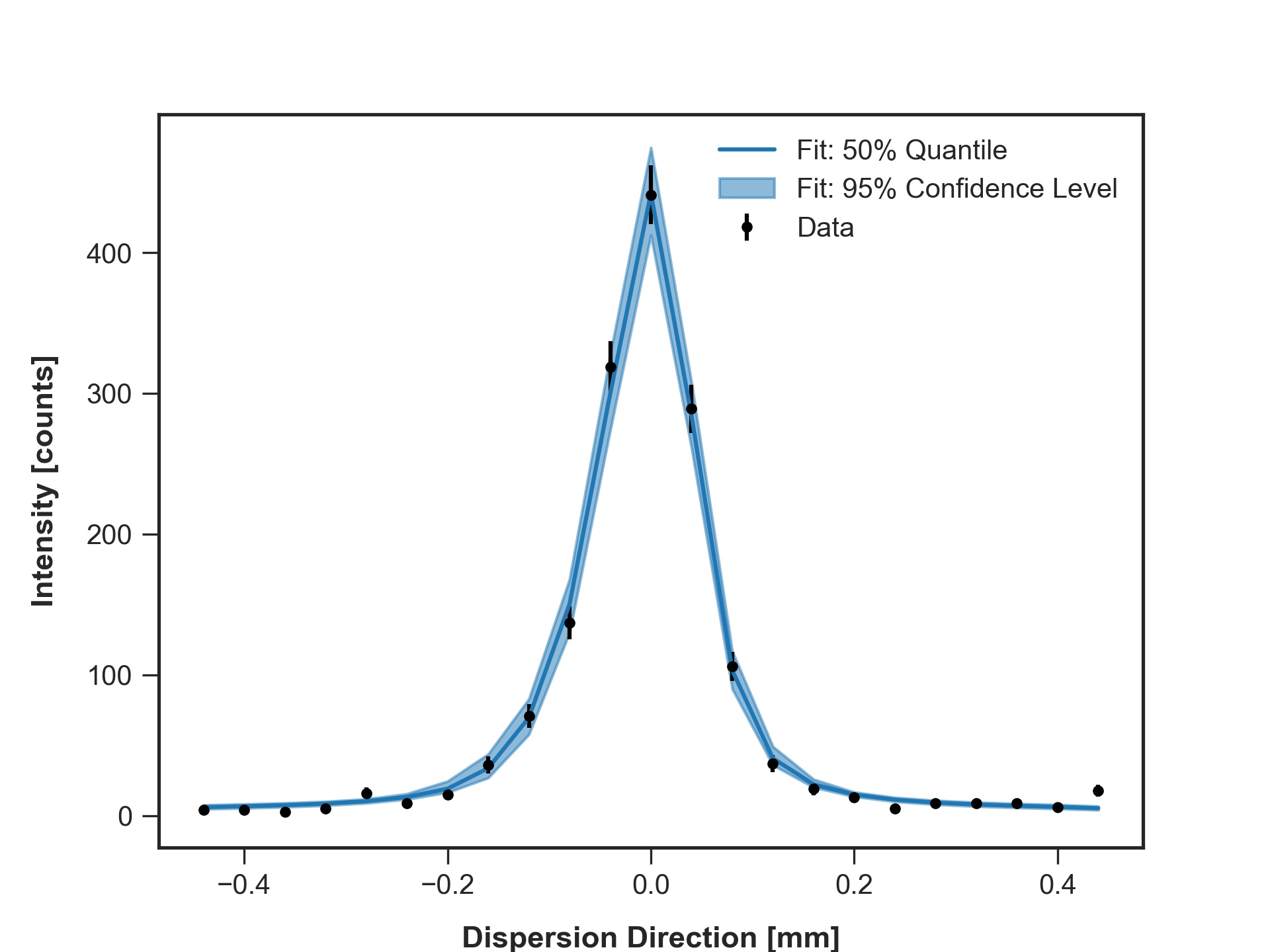}
\caption{The fitted fifth-order LSF profiles showing the 50\% quantile in dark blue with the 95\% confidence level filled in a lighter blue. The black data points represent the observed fifth-order LSF profile with their error bars described by Poisson statistics.}
\label{fig:fitted_lsf_profiles}
\end{figure*}

\begin{figure*}
\centering
\includegraphics[width=0.5\linewidth]{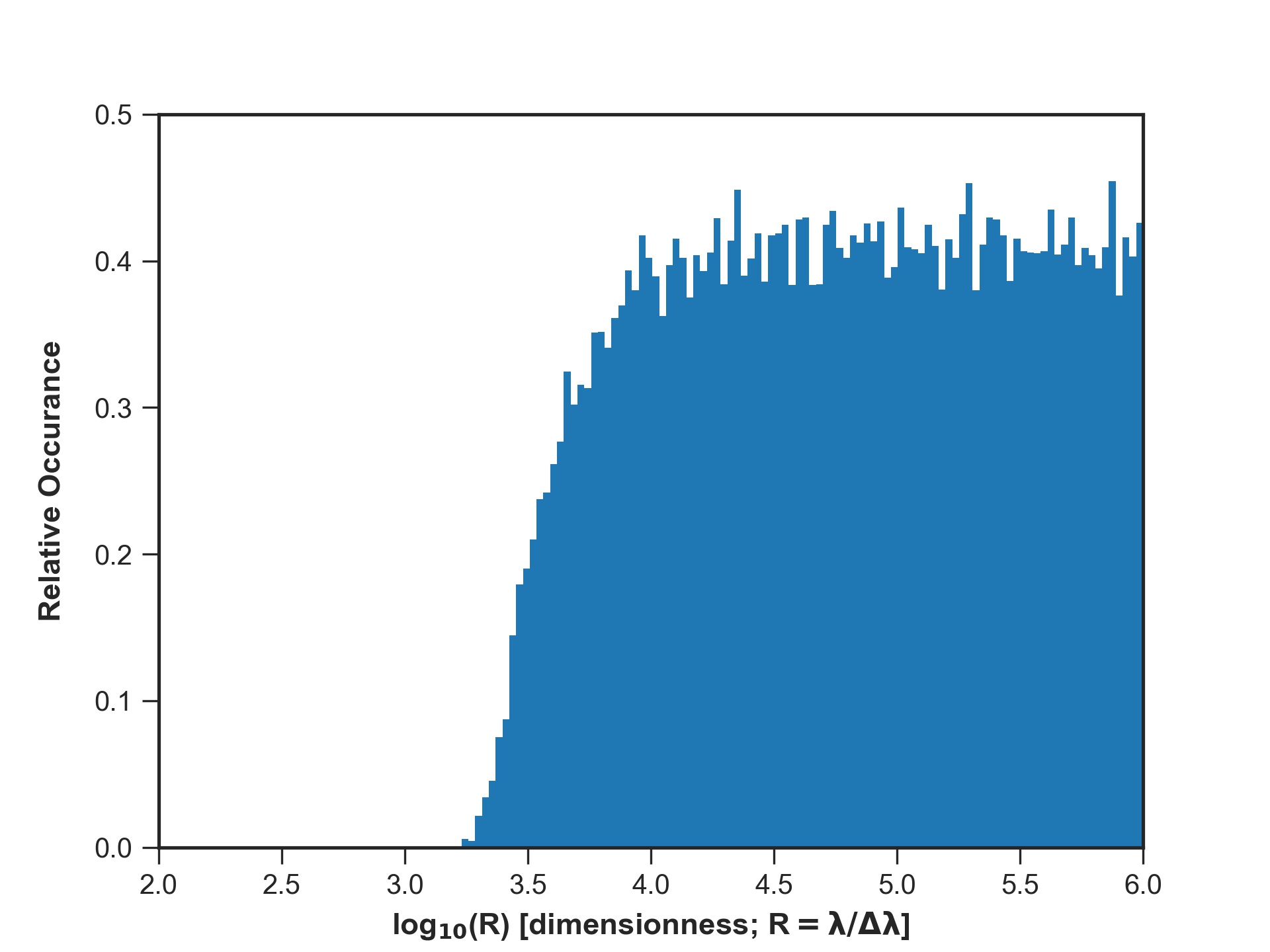}
\caption{The posterior probability distribution of $\log(R_g)$ across the sampled parameter space of $\log(R_g)=2-6$. The prior probability distribution of this parameter was a Uniform distribution from $\log(R_g)=2-6$.}
\label{fig:res_posterior}
\end{figure*}

The MCMC sampler was then run again with the parameters recommended by the Raftery-Lewis Diagnostic ($N_{step}=2,010,000$, $N_{burn}=10,000$, $N_{thin}=40$) to produce a $50,000$ step sample of the posterior probability distribution. The resulting 2.5\%, 50\%, and 97.5\% quantile levels of the best-fit model are presented in Figure \ref{fig:fitted_lsf_profiles}. Further, the posterior probability distribution for the parameter of interest, $\log(R_g)$, is shown in Figure \ref{fig:res_posterior}. As can be seen in this figure, the posterior probability distribution of $\log(R_g)$ disfavors values below $\log(R_g)\approx3.3$ and increasingly favors higher $\log(R_g)$ values, plateauing in ``favorability'' at $\log(R_g)\approx4$. 
%The Raftery-Lewis Diagnostic was then run on this sample to confirm the previous Raftery-Lewis Diagnostic results. This second result was consistent with the initial $N=10,000$ step sample result, confirming the initial Raftery-Lewis Diagnostic result and indicating that the final chain had converged.

\section{Discussion} \label{sect:discussion}

In this test campaign, a reflection grating prototype for the OGRE soft X-ray spectrometer was tested at the PANTER X-ray Test Facility to evaluate its performance. The results presented in Section \ref{sect:results} show that the model disfavors diffraction-induced resolutions below $\log(R_g)\approx3.3$ and increasingly favors resolutions above this lower limit, plateauing in favorability at $\log(R_g)\approx4$. Upper limits on the $\log(R_g)$ posterior are consistent with the upper limit placed on the prior distribution -- $\log(R_g)\sim6$. This result is first compared with the measurement capability of the assembled spectrometer. Then, the result will be put into the context of the OGRE single-grating resolution requirement. 

\subsection{Measurement Capability}

To understand whether the obtained $\log(R_g)$ posterior probability distribution is due to diffraction-induced errors or can be described by a fundamental limit in the measurement capability of the system, a raytrace simulation similar to that performed in Section \ref{sect:optic_zero_comp} was constructed to simulate the observed fifth-order LSF. If the $\log(R_g)$ posterior probability distribution truly describes the diffraction-induced errors in the system, then an additional blur term (consistent with $\log(R_g)>3.3$) will be required to best simulate the observed fifth-order LSF. However, if the $\log(R_g)$ posterior probability distribution does not describe these diffraction-induced errors, then no additional blur will be needed. 

\begin{figure*}
    \centering
    \begin{subfigure}[ht]{0.48\linewidth}
        \includegraphics[width=\linewidth]{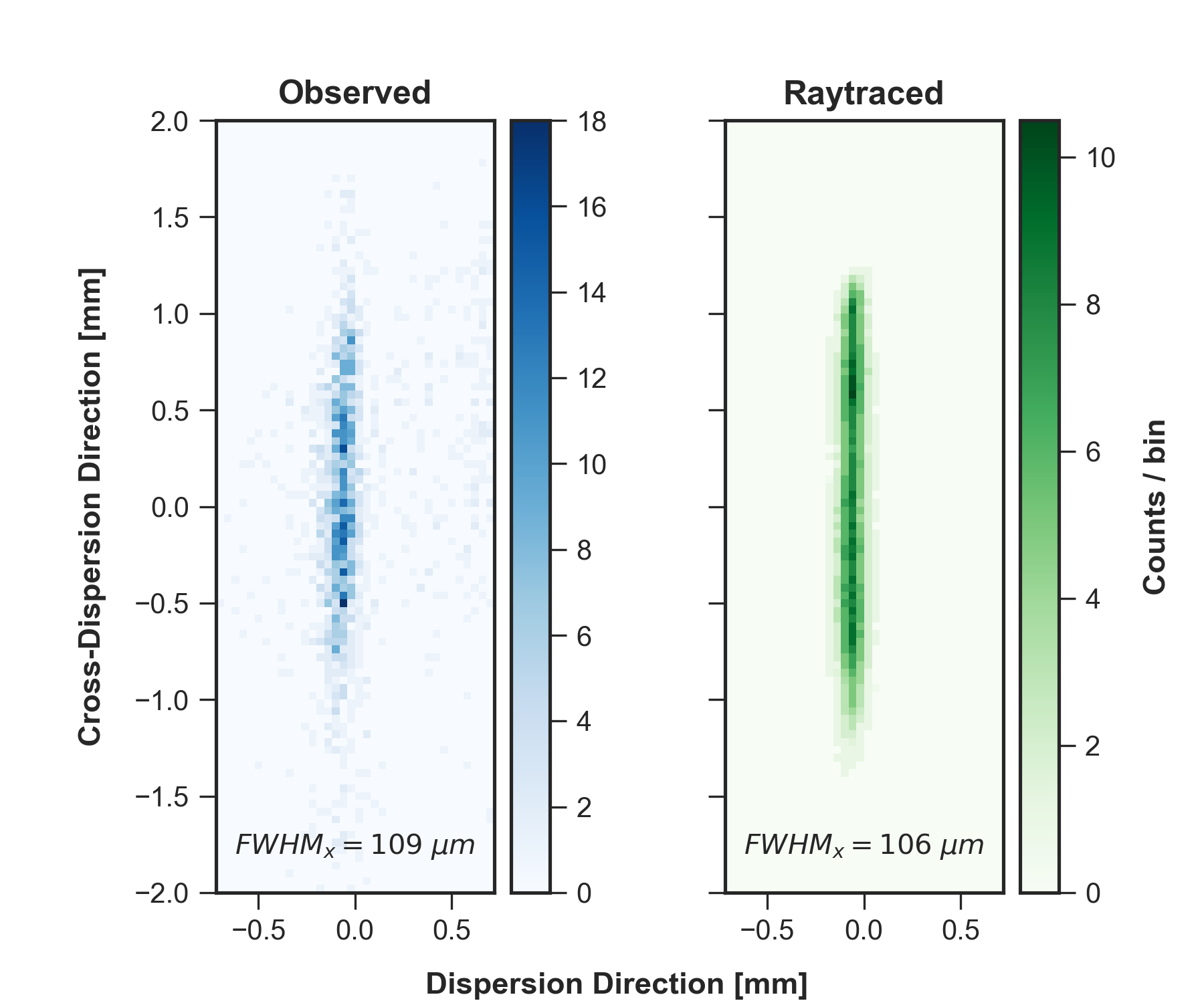}
        \caption{}
    \end{subfigure}
    \hfill
    \begin{subfigure}[ht]{0.48\linewidth}
        \includegraphics[width=\linewidth]{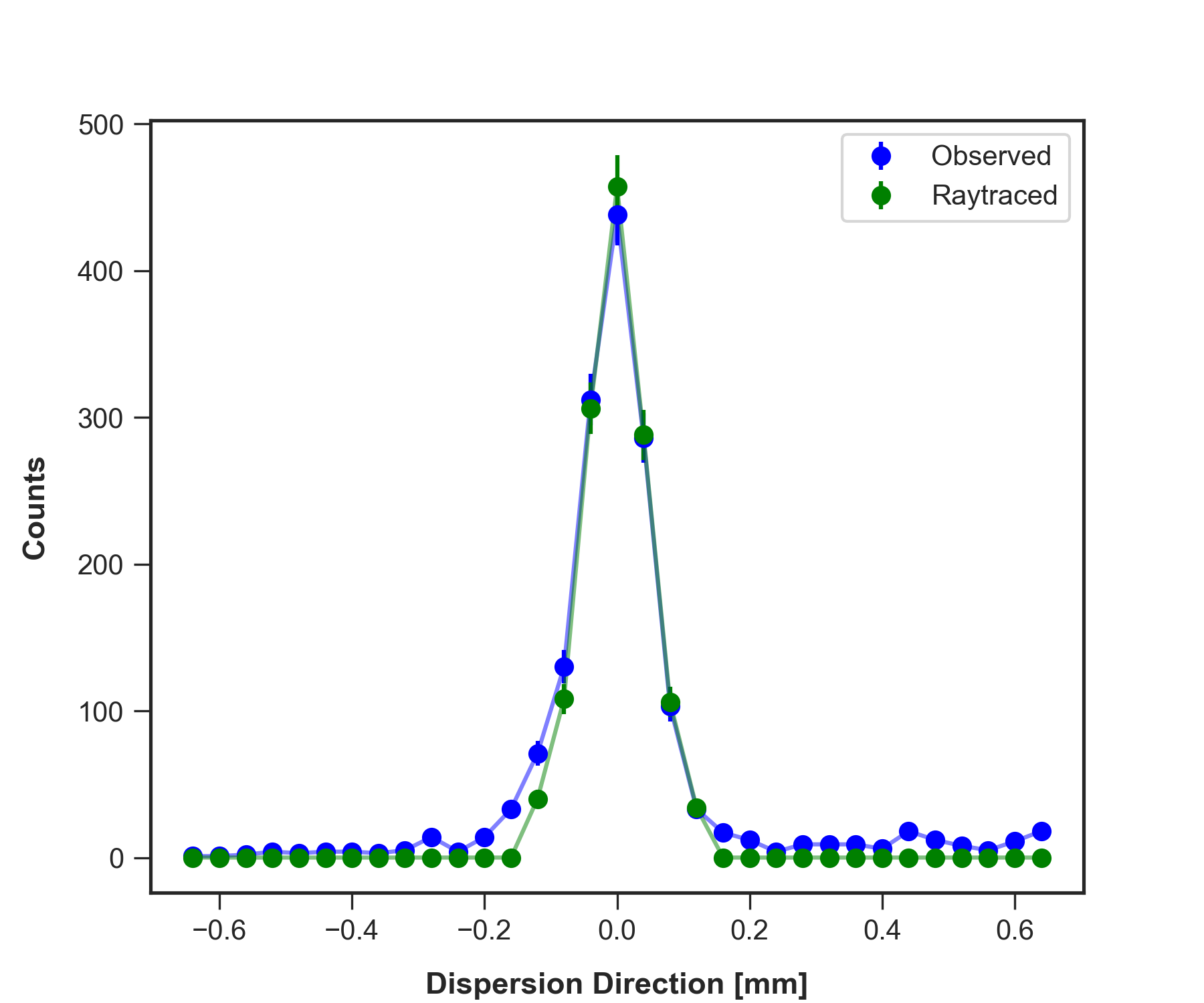}
        \caption{}
    \end{subfigure} 
\caption{A comparison of the observed (blue) and simulated (green) fifth-order LSF (a) and the resulting LSF profile when collapsed onto the dispersion direction (b).}
\label{fig:sim_fifth_order}
\end{figure*}

To simulate the observed diffracted-order LSF, a raytrace simulation similar to the simulation described in Section \ref{sect:optic_zero_comp} was constructed. An optic in accordance with the parameters described in Section \ref{sect:optic} was first simulated, and the resulting rays were then reflected off of a grating with a figure consistent with that shown in Figure \ref{fig:grat_fig}. Beckmann and Gaussian scatter were then systematically added to the simulated rays to reproduce the shape of the observed zero-order LSF on the focal plane. These rays, which now accurately reproduce the zero-order LSF on the focal plane, were then diffracted off of the grating and propagated to the focal plane. The wavelength of each ray was drawn from a double Lorentzian profile with line parameters consistent with those presented in Table \ref{tab:line_params}. No other effects, such as grating period errors or other diffraction-induced errors, were simulated. The results of the raytrace simulation are presented in Figure \ref{fig:sim_fifth_order}. The simulated fifth-order LSF profile had a FWHM of $106\substack{+14 \\ -12}$ \si{\micro\metre}. This width is consistent with the observed fifth-order LSF width of $109\substack{+11 \\ -13}$ \si{\micro\metre} FWHM well within $2\sigma$ errors. The results of this simulation indicate that the raytrace simulation of the fifth-order LSF profile does not need any additional width, such as that from diffraction-induced errors ($R_g$), to be consistent with the observed fifth-order LSF profile. 

% Before simulating the diffracted-order LSF though, the raytrace simulation from Section \ref{sect:optic_zero_comp} was first adjusted to better represent the observed zero-order LSF profile. In Section \ref{sect:optic_zero_comp}, an optic with Beckmann and Gaussian scatter was simulated to determine whether the observed optic PSF could be reproduced. However, the s These rays were then reflected off of the grating figure in Figure \ref{fig:grat_fig}. While this simulation agreed with the observed optic PSF, it did not produce a zero-order LSF profile that agreed with the observed zero-order LSF (observed: $76\substack{+4 \\ -5}$ \si{\micro\metre} FWHM; simulated: $100\substack{+7 \\ -9}$ \si{\micro\metre} FWHM). To make the simulated zero-order profile consistent with the observed data, the Gaussian scatter used to simulate the optic was reduced until the simulated zero-order profile agreed with the observed zero-order profile.

% $100\substack{+7 \\ -9}$ \si{\micro\metre} FWHM in the dispersion direction. This width does not agree with the observed zero-order LSF width of $76\substack{+4 \\ -5}$ \si{\micro\metre}

\begin{figure*}
\centering
\includegraphics[width=0.5\linewidth]{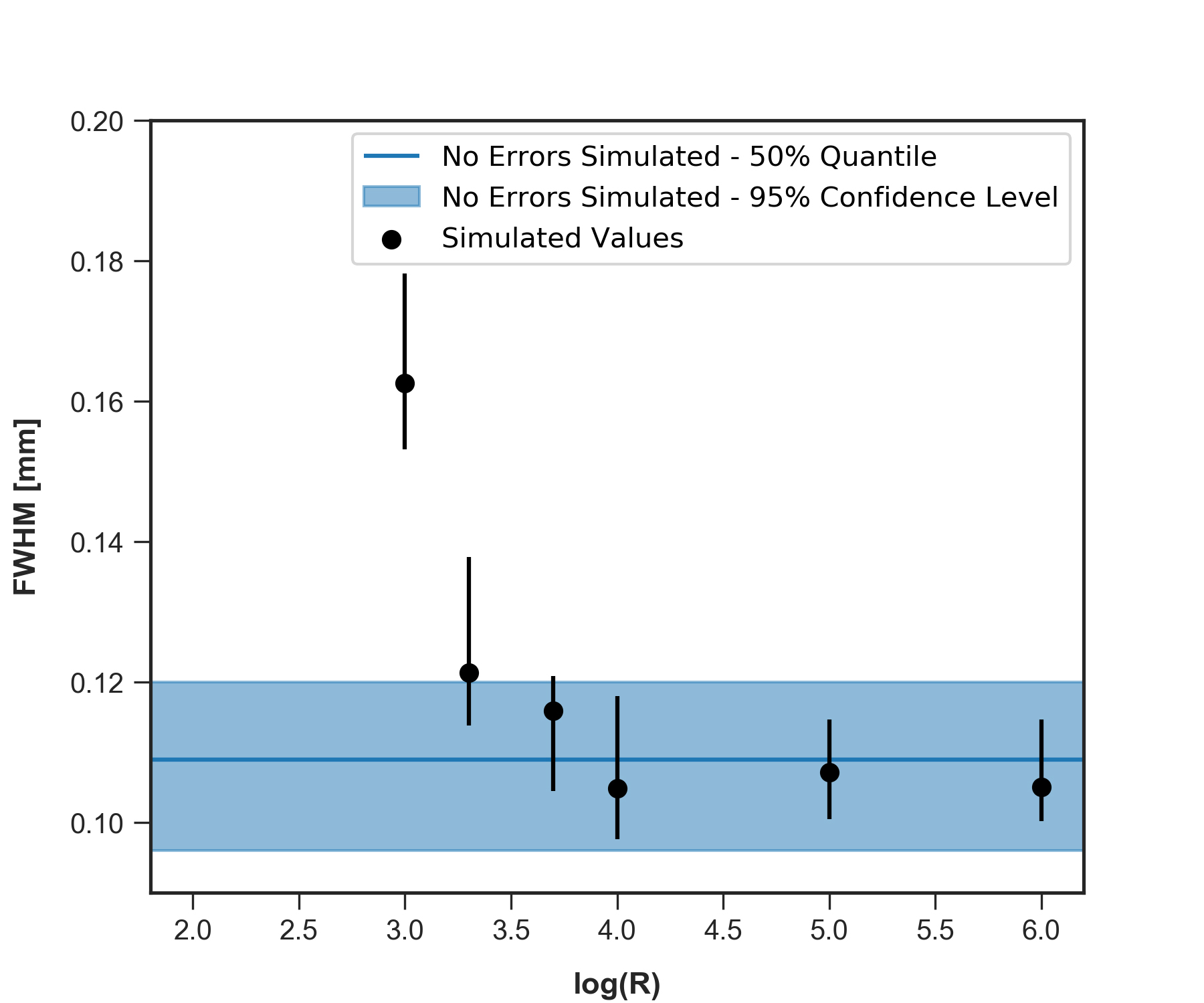}
\caption{Results from raytrace simulations of the tested system with various amounts of diffraction-induced errors (represented by $\log(R_g)$). This plot shows that increasing $\log(R_g)$ values have an increasing level of agreement with the results of a raytrace simulation without diffraction-induced errors (blue). A data point at $\log(R_g)=2.0$ was simulated, but is at a FWHM of $>0.8$ mm.}
\label{fig:grat_res_sim}
\end{figure*}

This raytrace simulation was further extended by simulating diffraction-induced errors over the modeled parameter space of $\log(R_g)=2-6$, and comparing these simulation results to the result without any diffraction-induced errors. The results of these simulations are presented in Figure \ref{fig:grat_res_sim}. For each data point, a Gaussian diffraction-induced error component with a width consistent with $R_g=x/2.355\sigma$ (with $x=102$ mm) was added to the raytrace simulation presented in the previous paragraph. The resulting width of the fifth-order LSF was then calculated for each simulated $R_g$ value. Figure \ref{fig:grat_res_sim} shows that higher $R_g$ values show an increasing level of agreement with the no diffraction-induced error case from Figure \ref{fig:sim_fifth_order} (equivalently, $\log(R_g)=\infty$). Diffraction-induced errors of $\log(R_g)\lesssim3.3$ are inconsistent with the $\log(R_g)=\infty$ case, but errors beyond this level become increasingly consistent with this case. Above $\log(R_g)\sim4$, all simulations are consistent with $\log(R_g)=\infty$. This behavior is similar to what is observed in the $\log(R_g)$ posterior probability distribution presented in Figure \ref{fig:res_posterior} and points to a measurement limit of $\log(R_g)\sim3.3-4$ for the as-tested system with the collected data. 

% This suggests that only the figure of the grating substrate, the performance of the optic, and geometric aberrations from diffraction are needed to reproduce the observed fifth-order LSF profile.

% As an additional check on the results of this raytrace, the results in Figure \ref{fig:res_posterior} are compared to the width of the observed fifth-order LSF profile -- $109\substack{+11 \\ -13}$ \si{\micro\metre} FWHM. The capability to measure additional contributions is limited by the error on the width of the observed LSF profile. Any errors that contribute at a level such that the LSF still has a width of $94-120$ \si{\micro\metre} FWHM cannot be measured accurately, as this result would still be consistent with the observed fifth-order LSF. 

As an additional check on this result, the measurement limit of the system was also calculated using the knowledge on the observed fifth-order LSF profile width: $109\substack{+11 \\ -13}$ \si{\micro\metre} FWHM. The capability to measure additional contributions is limited by the uncertainty on the width of the observed LSF profile. Any contributions that contribute such that the observed LSF still has a width of $<120$ \si{\micro\metre} FWHM cannot be measured accurately, as this result would be consistent with the observed profile within $2\sigma$ errors. Assuming that errors to the LSF add in quadrature, an error would need to contribute at the $\sim50$ \si{\micro\metre} level to push the observed LSF width past its $2\sigma$ confidence level. This $\sim50$ \si{\micro\metre} error is equivalent to $\log(R_g)\approx3.3$ at a dispersion of $\sim102$ mm, which agrees well with the lower limit of the $\log(R_g)$ posterior probability distribution in Figure \ref{fig:res_posterior}. This also agrees with the results of the raytrace simulations presented in Figure \ref{fig:grat_res_sim}. Any spatial errors lower than this $\sim50$ \si{\micro\metre} limit, or equivalently resolution errors greater than this $\log(R_g)\approx3.3$ limit, would be increasingly consistent with the observed data until a limit is reached where these contributions are indistinguishable. This expected behavior is also consistent with what is observed in Figure \ref{fig:res_posterior} and in Figure \ref{fig:grat_res_sim}. Thus, it is posited that the resulting $\log(R_g)$ posterior probability distribution can be explained solely by the measurement limit of the system and not diffraction-induced errors based on the results presented in this subsection. The system possesses varying levels of sensitivity below $\log(R_g)\approx4$, but cannot discern between errors beyond this limit.

\subsection{OGRE Summary}

The single-grating resolution requirement for the OGRE payload is $R_g>4500$ \citep{Donovan:2020aa}, meaning that each grating in the spectrometer must perform at this level to achieve the performance goals of OGRE. As outlined in Section \ref{subsect:grat_res_requirement}, this requirement flows from a comprehensive LSF error budget which delineates potential errors that affect the LSF in the spectrometer and allocates permissible values for each of these errors. The errors combine to form the final observed shape of the OGRE LSF, and therefore must be constrained to meet the performance goals of the instrument. Placing the test results into the context of the OGRE payload, Figure \ref{fig:res_posterior} shows that many of the posterior samples are consistent with this resolution value or higher. For the parameter space of $\log(R_g)=2-6$ that was sampled, the OGRE grating prototype achieved $R_g>4500$ at the $\sim94\%$ confidence level. This result is likely a conservative estimate though, since the $\log(R_g)$ posterior probability distribution in Figure \ref{fig:res_posterior} is skewed heavily due to the finite measurement limit of the system. The true diffraction-induced aberrations may be consistent with values much higher than $\log(R_g)=3.3-4$, but they cannot be measured with the system presented in this manuscript.

\subsection{Future Work}

The assembled spectrometer increasingly cannot characterize diffraction-induced errors above $R_g\sim2000$ ($\equiv\log(R_g)=3.3$) and becomes completely unable to discern between diffraction-induced errors above $R_g\sim10,000$  ($\equiv\log(R_g)=4$). To better probe spectral resolutions beyond these limits, improvements could be made to the system in the following three areas: (1) increased integration times, (2) improved grating figure, and (3) improved optic PSF extent in the dispersion direction. The most straight-forward improvement to the system is obtaining more data through increased integration times, as this improvement does not rely on changing anything in the system other than data acquisition time. Increased integration times reduce the uncertainties in the observed zero-order and fifth-order LSF profiles, thereby improving the knowledge of these profiles and allowing the model to better discern between different contributors to the diffracted-order LSF. For example, if the uncertainty in the width of the observed fifth-order LSF profile could be reduced from $\sim11$ \si{\micro\metre} to $\sim2$ \si{\micro\metre}, the assembled system could have characterized diffraction-induced aberrations at the $R\approx4800$ level -- beyond the single-grating resolution requirement of $R>4500$ that is required for the OGRE payload. Integration times can quickly become impractical though as integration time scales as roughly $\sigma^{-2}$.

% Estimates indicate that an improvement of this magnitude would necessitate a factor of $\sim30$x increase in integration time. For this test campaign, this would equate to integration times for the final exposures of the zero-order and fifth-order LSFs of $\sim13$ hours. 

\begin{figure}
\centering
\includegraphics[width=0.5\linewidth]{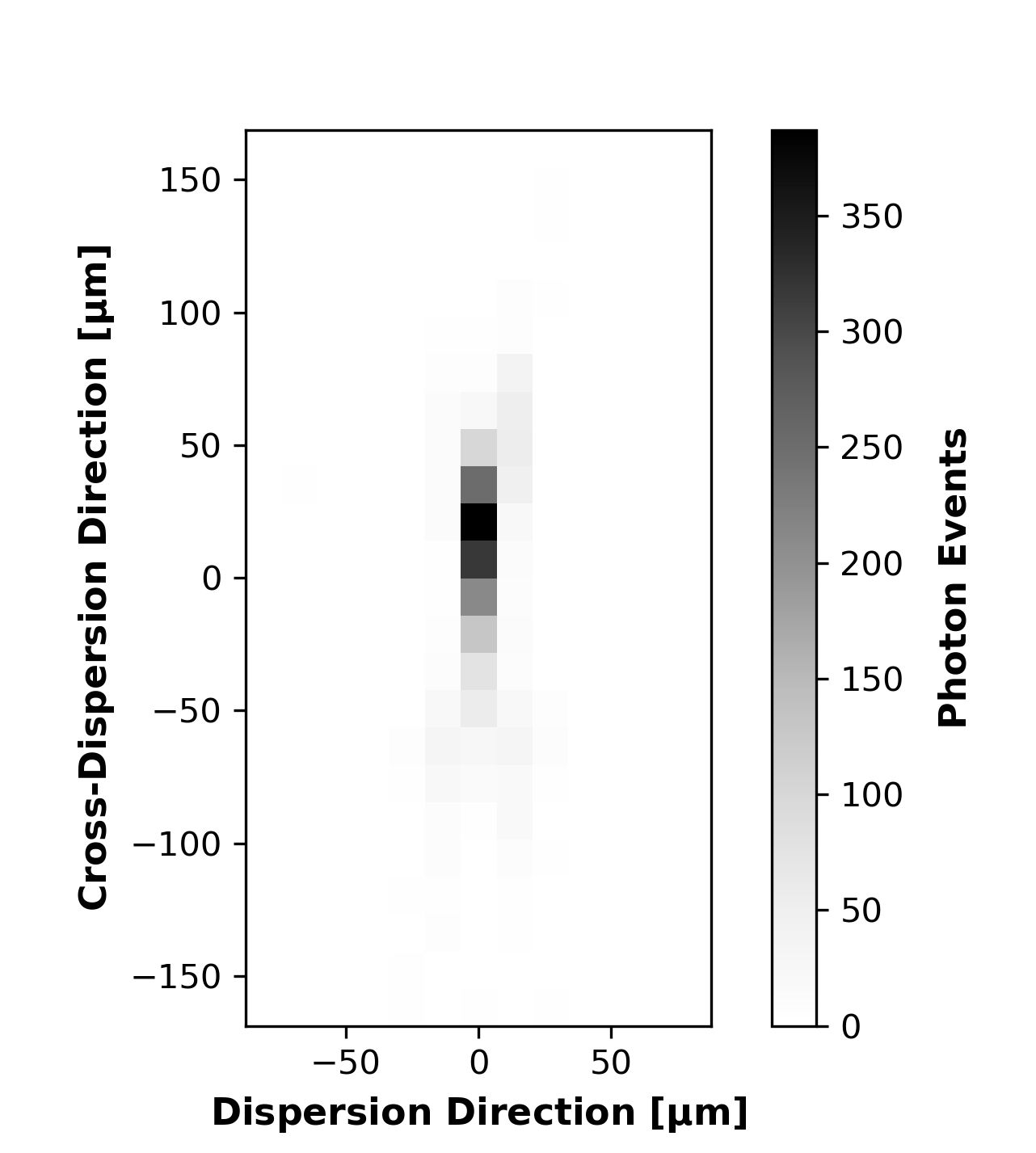}
\caption{Preliminary measurement of an single-shell OGRE optic prototype manufactured by NASA GSFC. The width in the dispersion direction is $<13.5$ \si{\micro\metre} FWHM and is limited currently by the pixel pitch of the detector in the test setup (13.5 \si{\micro\metre}). The width in the cross-dispersion direction is $\sim1.5''$ ($ \approx28$ \si{\micro\metre}) HPD.}
\label{fig:ogre_optic_results}
\end{figure}

In lieu of increased integration times, improving the figure of the grating substrate and/or the dispersion extent of the optic PSF that is feeding the spectrometer could also increase the achievable system resolution and therefore allow future tests to probe increasing diffraction-induced aberrations. A single-shell OGRE optic prototype has recently demonstrated a cross-dispersion extent of $\sim1.5''$ HPD and a dispersion extent of $<13.5$ \si{\micro\metre} FWHM during testing at Penn State's X-ray test facility. The X-ray source in this facility has an angular size of $\sim0.5''$ ($=9.2$ \si{\micro\metre} on the focal plane), which results in an optic performance of $\sim1.4''$ HPD in the cross-dispersion dimension and $<9.8$ \si{\micro\metre} FWHM in the dispersion dimension. An image of the PSF formed by this optic in this facility is presented in Figure \ref{fig:ogre_optic_results}. Furthermore, grating prototypes are now being produced on much flatter substrates. Rather than grating substrates that have a peak-to-valley of $\sim18$ \si{\micro\metre}, new substrates are routinely being manufactured on grating substrates with a peak-to-valley of $<2$ \si{\micro\metre}. Future tests with these two improvements would be able to realize observed zero-order LSF profiles with widths on the order of $<14$ \si{\micro\metre} FWHM -- a factor of $\sim5$x improvement over what was obtained in this test campaign. At the diffraction geometry presented in Table \ref{tab:diff_geom}, this equates to system-limited spectral resolutions of $R>7200$. By implementing some combination of these three discussed improvements, future tests will be able to probe spectral resolutions far beyond what was achieved in this test campaign.

\section{Summary}
In this work, a X-ray reflection grating spectrometer was assembled at the PANTER X-ray Test Facility and tested for performance. The spectrometer was composed of a mono-crystalline silicon X-ray optic from NASA GSFC, an X-ray reflection grating manufactured via e-beam lithography at The Pennsylvania State University's Materials Research Institute, and a detector at the focal plane. The grating was the first grating prototype manufactured for the OGRE soft X-ray spectrometer. The grating was tested in an OGRE-like diffraction geometry (Table \ref{tab:diff_geom}) to compare its measured performance to the performance required as outlined in the OGRE comprehensive LSF error budget \cite{Donovan:2020aa}. 

A Bayesian model was constructed for the observed diffracted-order LSF profile. This model was composed of a model for the observed zero-order LSF profile, a theoretical model for the Mg-K$\alpha$ doublet line profile that was used to test the spectrometer, and an unknown diffraction-induced error component. A MCMC sampler was used to sample the posterior probability distribution of the model and the unknown diffraction-induced error component. Results indicate that the tested grating achieved the OGRE single-grating resolution limit of $R_{g}>4500$ at the 94\% confidence level. Further, the posterior probability distribution of $\log(R_g)$ disfavors all resolution values below $R_{g}\sim2000$ and increasingly favors higher resolution values. It is, however, posited that the obtained posterior probability distribution is not dominated by diffraction-induced errors, but instead due to the measurement limit of the assembled spectrometer. Therefore, these results would be a conservative lower limit for diffraction-induced errors. Raytrace simulations of the diffracted-order LSF with realistic modeling of the optic, grating figure, and spectral line profile indicate that no diffraction-induced errors are needed to explain the observed diffracted-order LSF profile. 

To further probe the diffraction-induced errors of OGRE grating prototypes, future tests should utilize some combination of increased integration times, better grating figure, and an increased knowledge of the relative X-ray source and optic contributions to the optic PSF. The PSF of a recently built single-shell OGRE optic prototype tested at Penn State's X-ray test facility with a $\sim0.5''$ X-ray source improves upon the PSF formed by the optic prototype and X-ray source used in this test campaign. Further, reflection gratings are now being manufactured on much flatter substrates to improve figure-induced aberrations. Finally, testing in the dedicated X-ray testing facility at Penn State will allow increased integration times to be obtained. The implementation of some, or all, of these improvements will allow future reflection grating tests to probe spectral resolutions beyond what has been achieved for this test campaign.

\section*{Acknowledgments}
The presented work was supported by NASA grant NNX17AD19G, internal funding from The Pennsylvania State University, and a NASA Space Technology Research Fellowship. Grating fabrication was carried out at the Materials Research Institute at The Pennsylvania State University. The authors would like to thank the staff at the Materials Research Institute for their assistance with grating fabrication; Bernd Budau, Andreas Langmeier, and Stefan Pa{\ss}lack at the PANTER X-ray Test Facility and Veronika Stehl\'ikov\'a from the Max Planck Institute for Extraterrestrial Physics for supporting this test campaign; and William W. Zhang and his research group at NASA Goddard Space Flight Center for providing the optic used for this test campaign. This work makes use of PyXFocus, an open source Python-based raytracing package.

\bibliographystyle{ws-jai}
\bibliography{sample}

\end{document}